\documentclass[usenatbib,hyphens]{mnras}

\usepackage{amssymb}
\usepackage{amsmath}
\usepackage{natbib}
\usepackage[pdftex]{graphicx}
\usepackage[normalem]{ulem}
\usepackage{mathtools}
\usepackage{cuted}
\usepackage{xfrac}
\usepackage[flushleft]{threeparttable}

\title[Fullerene evolution in the ISM]
{Efficiency of the top-down PAH-to-fullerene conversion in UV irradiated environments}
\author[M. S. Murga et al.]
{M. S. Murga$^{1,2}$\thanks{E-mail: murga@inasan.ru},
 V. V. Akimkin$^{1}$, D. S. Wiebe$^{1}$\\
$^{1}$Institute of Astronomy, Russian Academy of Sciences, Pyatnitskaya str. 48, Moscow 119017, Russia,\\
$^{2}$Faculty of Chemistry, Lomonosov Moscow State University, Universitetsky pr. 13, Moscow 119234, Russia}

\date{Accepted today. Received tomorrow; in original form \today}
\pubyear{2022}

\begin{document}
\label{firstpage}
\pagerange{\pageref{firstpage}--\pageref{lastpage}}\maketitle

\begin{abstract}
	Polycyclic aromatic hydrocarbons (PAHs) and fullerenes play a major role in the physics and chemistry of the interstellar medium. Based on a number of recent experimental and theoretical investigations we developed a model in which PAHs are subject to photo-dissociation (carbon and hydrogen loss) and hydrogenation. We take into account that dehydrogenated PAHs may fold into closed structures -- fullerenes. Fullerenes, in their turn, can be also hydrogenated, becoming fulleranes, and photo-dissociated, losing carbon and hydrogen atoms. The carbon loss leads to shrinking of fullerene cages to smaller ones. We calculate the abundance of PAHs and fullerenes of different sizes and hydrogenation level depending on external conditions: the gas temperature, intensity of radiation field, number density of hydrogen atoms, carbon atoms, and electrons. We highlight the conditions, which are favourable for fullerene formation from PAHs, and we conclude that this mechanism works not only in H-poor environment but also at modest values of hydrogen density up to 10$^{4}$~cm$^{-3}$. We found that fulleranes can be formed in the ISM, although the fraction of carbon atoms locked in them can be maximum around 10$^{-9}$. We applied our model to two photo-dissociation regions, Orion Bar and NGC~7023. We compare our estimates of the fullerene abundance and synthetic band intensities in these objects with the observations and conclude that our model gives good results for the closest surroundings of ionising stars. We also demonstrate that additional fullerene formation channels should operate along with UV-induced formation to explain abundance of fullerenes far from UV sources.
 
\end{abstract}


\section{Introduction}

Observations reveal significant abundances of aromatic molecules in the interstellar medium~\citep{allamandola85, giard94, peeters02, draine07, galliano08}. Two most straightforward representatives of these compounds are polycyclic aromatic hydrocarbons (PAHs) and fullerenes. While the interstellar infrared (IR) spectra do contain characteristic features of aromatic species, specific PAHs have not yet been identified in the interstellar medium (ISM), existence of interstellar fullerenes, C$_{60}$, C$_{60}^{+}$, and C$_{70}$, was demonstrated by observations of the pertinent IR emission bands at 17.4 and 18.9$\mu$m~\citep{cami10, sellgren10} and absorption bands at 9577 and 9632~\AA{}~\citep{campbell15}. It is believed that large carbonaceous molecules such as PAHs and fullerenes are formed in carbon-rich asymptotic giant branch stars~\citep{frenklach89, cherchneff11, cherchneff12}. However, these particles evolve further as they travel through the ISM, due to ultraviolet (UV) irradiation, bombardment of highly energetic particles, etc., and their abundances change accordingly~\citep{hony01, galliano08, mic10_shock, boersma12, castellanos14, murga_orion} along with their ionisation state and molecular structure~\citep{allamandola99, montillaud13, andrews16, peeters17, sidhu22}.

One of the major factors in the carbonaceous particle evolution is their interaction with UV radiation. PAHs undergo photo-processing in the diffuse ISM and photo-dissociation regions (PDRs). In spite of their stability, they can lose hydrogen atoms and edged carbon atoms after absorption of a UV photon~\citep{jochims94, allain96, lepage01}. In a reverse process, hydrogen atoms can be re-attached~\citep{rauls08, goumans11}. Thus, PAH size and hydrogenation states are determined mainly by two factors, which are intensity of UV radiation field and H atom number density. It is believed that the diffuse ISM is mostly populated by compact PAHs with number of carbon atoms ($N_{\rm C}$) from $\approx30$ to 150, while the mean size of PAHs increases closer to a UV radiation source~\citep{andrews15, croiset16, murga_orion, knight21}. According to \cite{montillaud13, andrews16}, small PAHs ($N_{\rm C}\lesssim 50$) are mostly dehydrogenated in PDRs, medium-sized PAHs ($50\lesssim N_{\rm C}\lesssim 90$) can be dehydrogenated under certain conditions, and larger PAHs ($N_{\rm C}\gtrsim 90$) do not lose their hydrogen atoms and can even be super-hydrogenated. Until PAHs have lost all their hydrogen atoms, the dissociation channel of H loss is more preferable after photon absorption because the binding energy of C-H bonds is smaller than that of C-C bonds. Once PAHs have become dehydrogenated, they start losing carbon atoms. Carbon loss may lead to formation of defective pentagon rings, bending of the molecule plane, and subsequent formation of fullerenes~\citep{chuvilin10, berne12, zhen14_ful}. Discovery of fullerenes and their relation with PAHs in PDRs indicate that this process (`top-down' formation) does occur in objects with enhanced UV radiation field~\citep{castellanos14, berne15}. The top-down fullerene formation from other carbonaceous grains is also possible~\citep{garcia11, mic12}.  

Once formed, fullerenes can undergo the same evolutionary processes as PAHs, i.e. photo-dissociation and hydrogenation. Activation energies of fullerenes are quite high ($\approx7-11$~eV~\citep{gluch04}), where the highest energy belongs to C$_{60}$. Therefore, their destruction is not expected to be efficient except in the closest vicinity of UV sources. The carbon loss likely leads to shrinking of cages as it was described in \cite{berne15}. Fullerenes were shown to be highly reactive with hydrogen~\citep{petrie95}, hence appearance of hydrogenated fullerenes (so-called fulleranes) can be expected in the ISM. Fulleranes were suggested to be possible carries of the emission bands near 3.5~$\mu$m~\citep{webster92}, anomalous microwave emission in molecular clouds~\citep{iglesias05, iglesias06}, the UV bump at 2175~\AA{} and also diffuse interstellar bands~\citep{iglesias04}. Therefore, fulleranes are being searched both in the ISM and in meteorites. In spite of their potential importance, there are no confident detections of fulleranes, unlike pure fullerenes, although some fingerprints may indicate on their presence~\citep{heymann97, palotas20, sabbah22}.

To estimate abundance of different fullerenes in a UV-dominated environment, a detailed modelling is required that includes all relevant evolutionary processes both for PAHs and fullerenes. \cite{berne15} presented photo-chemical evolutionary model of PAHs and fullerenes and estimated abundance of C$_{60}$ in the PDR NGC~7023. In this work, we extend the model setup of \cite{berne15} by taking into account processes of hydrogenation and dehydrogenation, adding C$_{70}$, and calculating the ionisation state of fullerenes. Our goal is to establish the range of conditions favourable to fullerene formation from PAHs subject to UV irradiation, and to check whether PAH dissociation is a sufficient source of fullerenes in objects with the enhanced UV radiation.

\section{Model of evolution of PAHs and fullerenes}

\begin{figure*}
\includegraphics[width=0.95\textwidth]{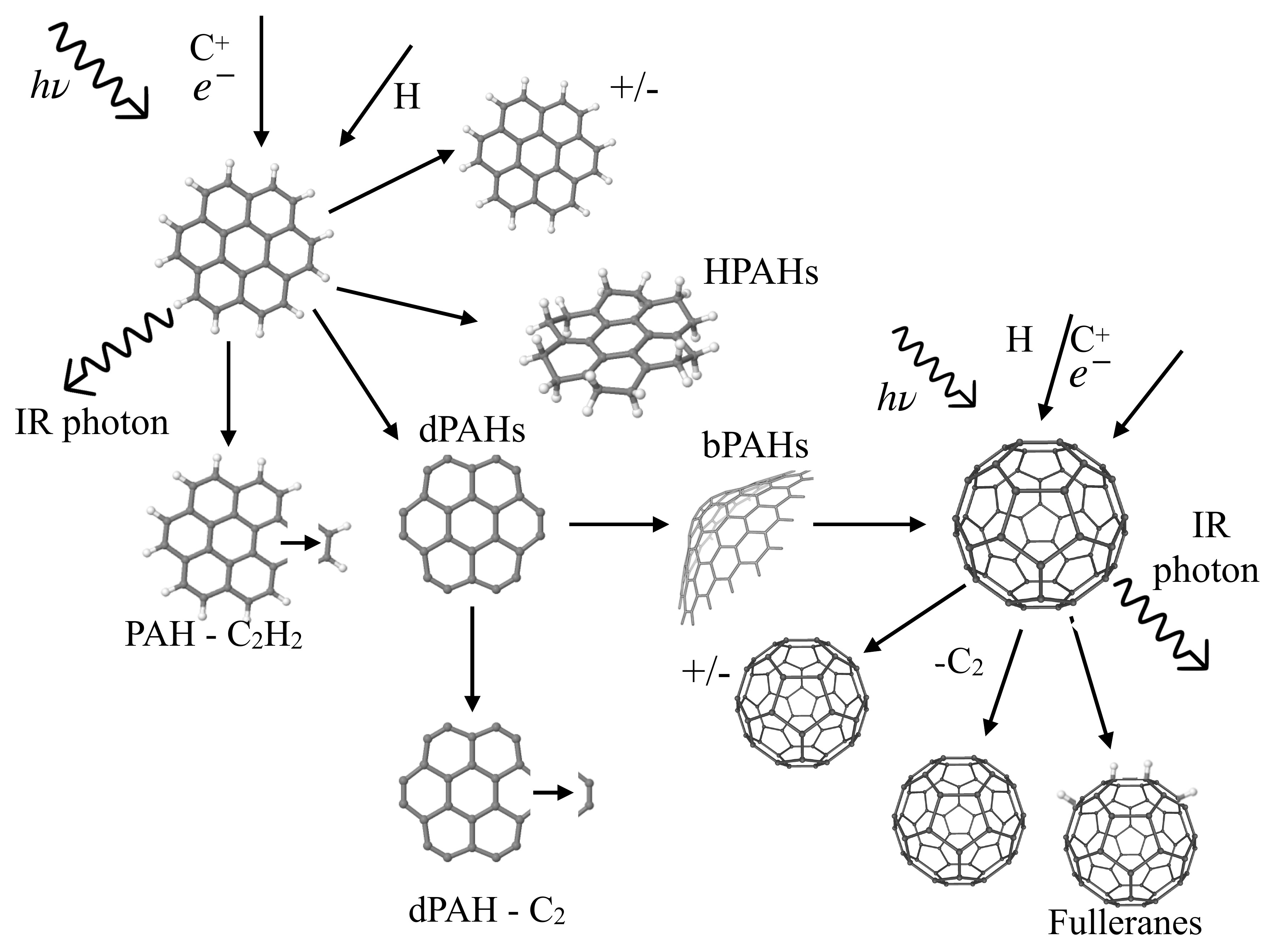}
\caption{Schematic illustration of our model with all the considered processes.}
\label{fig:scheme}
\end{figure*}

\begin{figure*}
\includegraphics[width=0.8\textwidth]{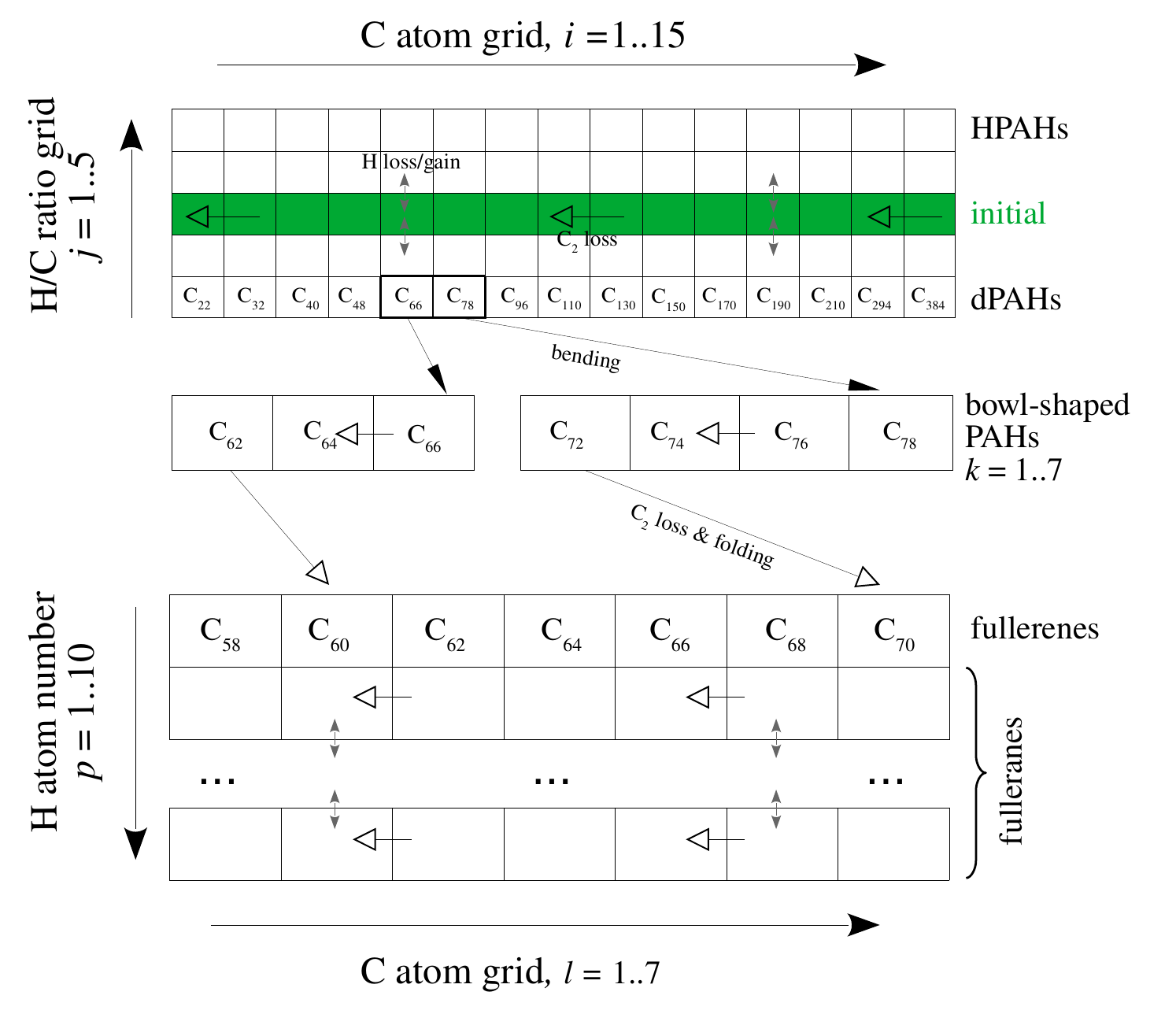}
\caption{Considered populations of PAHs, bowl-shaped PAHs and fullerenes and involved evolutionary processes~-- C$_2$ loss, (de)hydrogenization, dPAHs bending, C$_2$-loss-induced fullerene folding.}
\label{fig:schemebin}
\end{figure*}

To describe the evolution of PAHs and fullerenes, we use the system of kinetic equations similar to the one utilised in the {\tt Shiva} model~\citep{murga_shiva} with modifications relevant to the current task. Modifications regarding hydrogenation and dissociation of super-hydrogenated PAHs have been already described in \cite{murga_acet, murga_orion}. In this work we expand our model and consider isomerization of PAHs and subsequent fullerene formation. Overall, the following processes are included: photo-dissociation (C and H loss) and hydrogenation of PAHs, bending of dehydrogenated PAHs and folding into fullerene-like cages, photo-dissociation and hydrogenation of fullerene cages. Schematically, the model is illustrated in Fig.~\ref{fig:scheme}. Detailed information about the processes is given below. Here we briefly describe the main aspects of the model. Three basic classes of molecules are considered: PAHs, bowl-shaped PAHs (bPAHs), and fullerenes. PAHs are subdivided into regular PAHs, dehydrogenated PAHs (dPAHs), and super-hydrogenated PAHs (HPAHs). dPAHs are produced from PAHs by H atom loss due to photo-dissociation. Further, they transform to bPAHs and then to fullerenes with some probability or undergo skeleton dissociation, remaining in the PAHs category. For technical convenience, we describe bPAHs by equations separate from PAHs. There are separate equations for fullerenes as well.

We consider an initial set of PAHs consisting of $N_{\rm m}=15$ molecules: C$_{24}$H$_{12}$, C$_{32}$H$_{14}$, C$_{40}$H$_{16}$, C$_{48}$H$_{18}$, C$_{66}$H$_{20}$, C$_{78}$H$_{22}$, C$_{96}$H$_{24}$, C$_{110}$H$_{26}$, C$_{130}$H$_{28}$, C$_{150}$H$_{30}$, C$_{170}$H$_{32}$, C$_{190}$H$_{34}$, C$_{210}$H$_{36}$, C$_{294}$H$_{42}$, C$_{384}$H$_{48}$. The set is supposed to cover the whole range of PAH sizes, but is not meant to represent any specific family of PAHs. To describe evolution of the PAHs in mathematical terms, we define PAH size bins by the number of carbon atoms ($N_{\rm C}$, index $i$) and by the number of hydrogen atoms ($N_{\rm H}$, index $j$). In other terms, the bins are determined by total mass of carbon and hydrogen atoms in the molecule ($m_i$ and $m^{\rm H}_{ij}$, correspondingly). The number density of PAHs in each bin with corresponding $m_i$ and $m^{\rm H}_{ij}$ is designated as $N_{ij}$. The scheme illustrating the considered molecules and corresponding bins is presented in Fig.~\ref{fig:schemebin}.

PAHs undergo evolutionary migration from a bin to a bin, changing their carbon and hydrogen masses. The centres of carbon mass bins correspond to the masses of carbon atoms in molecules listed above. Hydrogenation or dehydrogenation of PAHs preserves their $i$ index, while changing $j$. In other words, the fraction of hydrogen atoms relative to carbon atoms ($X^{\rm H}_{ij}$) varies. We designate the ratio of $N_{\rm H}$ to $N_{\rm C}$ in the initial set as $X^{\rm H}_{ij,0}$. The range of hydrogen mass in PAHs is divided into $N_{\rm Hm}=5$ bins. All the PAHs from the initial set fall into bins with $j=3$. The bins with $j=1,2$ have $X^{\rm H}_{ij}$ equal to ($\sfrac{1}{3}X^{\rm H}_{ij,0}, \sfrac{2}{3} X^{\rm H}_{ij,0}$). The bins with $j=4,5$ have $X^{\rm H}_{ij}$ equal to ($1.5X^{\rm H}_{ij,0},  2 X^{\rm H}_{ij,0}$). Values of $m^{\rm H}_{ij}$ are calculated from $X^{\rm H}_{ij}$. Borders of the bins by carbon and hydrogen mass are designated as $m^{\rm b}_{i}$ and $m^{\rm Hb}_{ij}$, respectively, and are defined as the average values between $m_i$ of neighbouring bins with $i-1$ and $i+1$ in case of carbon and $m^{\rm H}_{ij}$ of neighbouring bins with $i,j-1$ and $i, j+1$ in case of hydrogen. The lower bounds $m^{\rm b}_{0}$ and $m^{\rm Hb}_{i0}$ are zero. The upper bounds $m^{\rm b}_{N_{\rm m}}$ and $m^{\rm b}_{N_{\rm Hm}}$ are determined as $m_{N_{\rm m}}+(m_{N_{\rm m}}-m_{N_{\rm m}-1}/2)$ and $m^{\rm H}_{iN_{\rm Hm}}+(m^{\rm H}_{iN_{\rm Hm}}-m^{\rm H}_{iN_{\rm Hm}-1}/2)$. The initial number density is calculated as $dn/dm(m^{\rm b}_{i+1}-m^{\rm b}_{i})$ with the mass distribution $dn/dm$ taken from \cite{wd01} (hereinafter WD01). 

We consider fullerene formation as a two step process. First, some dPAHs can start to bend and turn to bPAHs population. Second, bPAHs can further lose their carbon atoms completing the folding into fullerenes. The number density of bPAHs is designated as $N_k^{\rm b}$. In total, we consider $N_{\rm b}=7$ bPAHs ($k$ from 1 to 7): C$_{62}$, C$_{64}$, C$_{66}$, C$_{72}$, C$_{74}$, C$_{76}$, C$_{78}$. We assume that only dPAHs C$_{66}$ and C$_{78}$ ($i=5$ and $6$) can turn to bPAHs population (with $k=3$ and $7$, correspondingly). Certainly any dPAHs can undergo bending, but we do not consider them as we focus only on formation channels of fullerenes C$_{60}$ and C$_{70}$ and their respective bPAHs pre-cursors. Carbon loss in C$_{66}$ and C$_{78}$ leads to filling of the bins with $k=1,2$ and $k=4,5,6$, correspondingly. Note that not all dPAHs from $i=5,6$ bins turn to bPAHs. Some of them can undergo fragmentation without bending and stay in dPAHs population.

The number density of fullerenes is designated as $N^{\rm ful}_{lp}$. We consider $N_{\rm ful}=7$ fullerene molecules: C$_{58}$, C$_{60}$, C$_{62}$, C$_{64}$, C$_{66}$, C$_{68}$, C$_{70}$. The index $l$ points to the cage size. The index $p$ indicates the number of attached hydrogen atoms. The index varies from 1 to $N_{\rm Hm, ful}=10$, thus, the fullerenes can be hydrogenated and have up to nine hydrogen atoms\footnote{Certainly the more number of hydrogen can be attached to fullerenes. However, we limited the number by 9 atoms, hence we have 10 equations corresponding to different hydrogenation state of each fullerene. The more number of equations, the more model becomes time-consuming. Looking ahead we can conclude that this limitation does not affect the final results as fullerenes can be hardly hydrogenated up to 9 atoms, especially more than that. So, we suppose that our limitation is justified.}. When bPAHs C$_{62}$ and C$_{72}$ ($k=1$ and 4) lose their C$_2$ they transit to fullerene bins with $l=2$ and 7, respectively, herewith $p=1$. As the evolution goes, fullerene bins with $l=1,3-6$ can be populated. 

The system of kinetic equations for $N_{ij}$, $N^{\rm b}_k$ and $N^{\rm ful}_{lp}$ is 
\begin{equation} \label{eq: big formula}
\begin{split}
\begin{cases}
\frac{dN_{ij}}{dt} &= \underbrace{A_{ij+1}^{(1)}N_{ij+1} - A_{ij}^{(1)}N_{ij}}_{\text{H loss due to photo-dissociation}}\\
&+ \underbrace{B_{i+1j}N_{i+1j} - B_{ij}N_{ij}}_{\text{C loss due to photo-dissociation}}  \\
&+\underbrace{A^{(2)}_{ij-1}N_{ij-1} - A^{(2)}_{ij}N_{ij}}_{\text{H addition to PAHs}}\\
& -\underbrace{ \sum\limits_{k} I_{ijk}N_{ij}}_{\text{transition to bPAHs}}\\
\frac{dN^{\rm b}_k}{dt} &= \sum\limits_{ij} I_{ijk}N_{ij} + \underbrace{B_{k+1}N_{k+1}^{\rm b} - B_{k}N_{k}^{\rm b}}_{\text{C loss due to photo-dissociation}} \\
\frac{dN^{\rm ful}_{lp}}{dt} &= \underbrace{ \sum\limits_{k} F_{klp}N_{k}^{\rm b}}_{\text{transition to fullerenes}}\\
&+\underbrace{A^{(1)}_{lp+1}N_{lp+1}^{\rm ful} - A^{(1)}_{lp}N_{lp}^{\rm ful}}_{\text{H loss due to photo-dissociation}}\\
&+\underbrace{B_{l+1}N^{\rm ful}_{l+1p} -B_{l}N^{\rm ful}_{lp}}_{\text{C loss due to photo-dissociation}}\\
&+\underbrace{A^{(2)}_{lp-1}N_{lp-1}^{\rm ful} - A^{(2)}_{lp}N^{\rm ful}_{lp}}_{\text{H addition to fullerene cages}}\\
\end{cases}
\end{split}
\end{equation}
where $A_{\rm index}^{(1,2)}$ and $B_{\rm index}$ are rate coefficients of change the H and C mass, respectively, for the bins with indices $ij$, $k$ or $lp$ depending on a molecule type, $I_{ijk}$ is the rate coefficient of transition of $ij$th PAH bin into the $k$th bin of bPAHs due to PAH sheet bending, and $F_{klp}$ is the rate coefficient of the transition of the $k$th bin of bPAHs to the $lp$th fullerene bin. 

The expressions for $A_{ij}^{(1,2)}$ and $B_{ij}$ and their boundary values were given in \cite{murga_acet, murga_orion} (Eqs.~2, 4 and related Eqs.~3, 5, 6). The same expressions are used in this work although here we also calculate rates for bPAHs (indices $k$) and fullerenes ($lp$) instead of PAHs and indices $ij$. The calculations for all molecules are similar with only difference in parameters of reactions. The generalised expression for $A_{\rm index}^{(1,2)}$ is 
\begin{equation}
A_{\rm index}^{(1,2)} = \frac{\varepsilon_{\rm index}^{(1,2)}}{\Delta m^{\rm H}_{\rm index}},
\label{b1}
\end{equation}
where $\varepsilon_{\rm index}^{(1,2)}$ is the rate of H mass change, $\Delta m^{\rm H}_{\rm index}$ is the H mass interval of the bin which is $m^{\rm Hb}_{ij+1}-m^{\rm Hb}_{ij}$ for PAHs, and $m_{\rm H}$ for fullerenes. $A_{\rm index}^{(1)}$ is set 0 if $j=1$ or $j \geqslant N_{\rm Hm}$ and if $p=1$ or $p \geqslant N_{\rm Hm, ful}$. $A_{\rm index}^{(2)}$ is set 0 if $j \geqslant N_{\rm Hm}$ or  $p\geqslant N_{\rm Hm, ful}$. The rate $\varepsilon_{\rm index}^{(1)}$ and $\varepsilon_{\rm index}^{(2)}$ are expressed as 
\begin{eqnarray}
\varepsilon_{\rm index}^{(1)} = \frac{\mu_{\rm H}}{N_{\rm A}}R^{\rm index}_{\rm H} \nonumber \\
\varepsilon_{\rm index}^{(2)} = \frac{\mu_{\rm H}}{N_{\rm A}}H^{\rm index}, 
\end{eqnarray}
where $\mu_{\rm H}$ is the molar mass of hydrogen, $N_{\rm A}$ is the Avogadro number, and $R^{\rm index}_{\rm H}$ and $H^{\rm index}$ are the rates of detachment and attachment of H atoms to the molecule indicated by its index. The rates are described below in Sect.~\ref{photodestr} and \ref{hydro}. 

The generalised coefficient $B_{\rm index}$ is calculated as
\begin{equation}
B_{\rm index} = \frac{\mu_{\rm index}}{\Delta m_{\rm index}},
\label{b2}
\end{equation}
where $\mu_{\rm index}$ is the rate of C mass change, $\Delta m_{\rm index}$ is the C mass interval of the bin which is $m^{\rm b}_{i+1}-m^{\rm b}_{i}$ for PAHs, and $2m_{\rm C}$ for fullerenes and bPAHs. $B_{\rm index}$ is 0 when a) the index $i>N_{\rm m}$, b) the index $k>N_{\rm b}$, or c) $k=1,4$, d) the index $l=1,\;N_{\rm ful}+1$. The rate $\mu_{\rm index}$ can be found through 
\begin{equation}
\mu_{\rm index} = \frac{\mu_{\rm C}}{N_{\rm A}}R^{\rm index}_{\rm C},
\end{equation}
where $\mu_{\rm C}$ is the molar mass of C, and $R^{\rm index}_{\rm C}$ is the rate of detachment of C atoms from the molecule indicated by its index and are described below in Sect.~\ref{photodestr}. 

The rates $I_{ijk}$ and $F_{klp}$ can be estimated through the rate $R^{\rm index}_{\rm C}$ with appropriate parameters, which we discuss later. Herewith, in this case we do not consider the mass fraction that was transferred to a final bin because during the isomerization grains fully transit from bins with indices $ij$ or $k$ to bins with indices $k$ and $lp$, correspondingly. Only two dPAHs can isomerize in our treatment. Namely, we account for conversion of C$_{66}$ and C$_{78}$ ($j=1$ and $i=5,6$) to bPAHs C$_{66}$ ($k=3$) and C$_{78}$ ($k=7$), correspondingly. Thus, $I_{ijk}$ is non-zero if $ijk=5,1,3$ and $ijk=6,1,7$. As for transition of bPAHs to fullerenes, only bPAHs C$_{62}$ ($k=1$) and C$_{72}$ ($k=4$) isomerise to fullerenes C$_{60}$ ($lp=2,1$) and C$_{70}$ ($lp=7,1$), correspondingly. Thus, $F_{klp}$ is non-zero if $klp=1,2,1$ and $klp=4,7,1$. 
 
We limit our evolutionary model by the processes listed above, although there are many other processes that can be included and may play a non-negligible role in evolution of aromatic molecules. Among them such processes as addition of carbon atoms to PAHs and to fullerenes, formation of PAH-PAH, PAH-fullerene and fullerene-fullerene dyads and larger clusters can be mentioned. However, experimental and theoretical investigations of these processes are not complete and robust enough to describe them quantitatively. 

\subsection{Photo-dissociation}
\label{photodestr}

The rates of detachment of C and H atoms can be estimated as 
\begin{equation}
R^{\rm index}_{\rm C|H}= \int_{E_{0, {\rm index}}}^{E_{\rm max}} k_{\rm diss}({\rm index},E) \left(\frac{dp}{dE}\right)_{\rm index} dE,
\label{r_destr}
\end{equation}
where $E_{0, {\rm index}}$ is the activation energy, $E_{\rm max}$ is the upper limit of photon energy, $k_{\rm diss}(\rm{index}, E)$ is the dissociation rate of the parent molecule with the internal energy $E$, and $\left(\frac{dp}{dE}\right)_{\rm index}$ is the probability that the parent molecule has the internal energy between $E$ and $E+dE$. $E_{\rm a, index}$ is determined by a type of a molecule and reaction and is given below. We adopt $E_{\rm max}=40$~eV so that to cover the whole range of energies that considered molecules can have with taking into account their ability to accumulate photon energy due to the multi-photon mechanism. We estimate $dp/dE$ with stochastic heating mechanism described in \cite{guhathakurta89} and applied in our previous works~\citep{murga_shiva}. In the calculations of the probability function, it was already taken into account that a fraction of energy went to ionisation and infrared emission, i.e. $dp/dE$ provides the probability that this PAH will spend its energy exclusively to dissociation.

The dissociation is considered as a unimolecular reaction. There are several possible treatments to estimate the rate $k_{\rm diss}(\rm{index},E)$, described in detail by \cite{tielens05}. Although the different methods give results generally consistent with each other, in some cases the discrepancy may be up to 1--3 orders of magnitude. We studied the discrepancy between the methods and its influence on results of modelling separately in Murga et al. 2022 (in prep.). In this work, we chose the expression, which is based on the RRKM theory after inverse Laplace transformation (which we further refer to as the Laplace method). We refer interested readers to the works of \citet{leger89} and \citet[][Sect.~6.4]{tielens05} and works cited there for detailed description of the method and corresponding expressions. Here we discuss only the parameters that were put into the expressions depending on the molecule and the process type. These parameters are given in Table~\ref{par}. 

\begin{table*}
\caption{Adopted dissociation parameters $k_0$ and $E_0$.}
\label{par}
\begin{center}
\begin{threeparttable}
\begin{tabular}{|l|c|c|}
\hline
Grain type and fragment  &          $E_0$, eV & $k_0$, s$^{-1}$    \\
\hline
PAHs$^{-,0,+}$, $X_{\rm H}\leq X_{\rm H}^{0}$, H  &            4.3 & 1.4$\cdot 10^{16}$ (a, b)       \\
PAHs$^{-}$, $X_{\rm H}> X_{\rm H}^{0}$, H  &            $-2.5 \frac{X_{\rm H}}{X_{\rm H}^{0}}+6.8$ &  1.4$\cdot 10^{16}$     \\
PAHs$^{0}$, $X_{\rm H}> X_{\rm H}^{0}$, H  &           $-2.7 \frac{X_{\rm H}}{X_{\rm H}^{0}}+7$ & 1.4$\cdot 10^{16}$  \\
PAHs$^{+}$, $X_{\rm H}> X_{\rm H}^{0}$, H  &       $-2.4 \frac{X_{\rm H}}{X_{\rm H}^{0}}+6.7$ &  1.4$\cdot 10^{16}$     \\
PAHs$^{-, 0, +}$, $X_{\rm H}\leq X_{\rm H}^{0}$, H$_2$  &             --  & --     \\ 
PAHs$^{-,0,+}$, $X_{\rm H}\leq X_{\rm H}^{0}$, C$_2$H$_2$  &            4.6 & 1.4$\cdot 10^{16}$ (b)   \\
PAHs$^{-}$, $X_{\rm H}> X_{\rm H}^{0}$, C$_2$H$_2$   &           $-2.7 \frac{X_{\rm H}}{X_{\rm H}^{0}}+7.3$ & 1.4$\cdot 10^{16}$      \\
PAHs$^{0}$, $X_{\rm H}> X_{\rm H}^{0}$, C$_2$H$_2$  &          $-2.9 \frac{X_{\rm H}}{X_{\rm H}^{0}}+7.5$ & 1.4$\cdot 10^{16}$     \\
PAHs$^{+}$, $X_{\rm H}> X_{\rm H}^{0}$, C$_2$H$_2$    &            $-2.6 \frac{X_{\rm H}}{X_{\rm H}^{0}}+7.2$ & 1.4$\cdot 10^{16}$       \\
dPAHs, folding      &   4.0 & 1.0$\cdot 10^{15}$ (c)   \\
bPAHs, C$_2$      &   4.6 & 1.4$\cdot 10^{16}$    \\
C$_{58}$(C$_{58}^{+}$), C$_2$   &  9.6 (9.8) & 5.0$\cdot 10^{19}$ (d, e) \\
C$_{60}$(C$_{60}^{+}$), C$_2$   &  11.2 (10.6) & 5.0$\cdot 10^{19}$ (d, e) \\
C$_{62}$(C$_{62}^{+}$), C$_2$   &  8.0 (8.5) & 5.0$\cdot 10^{19}$ (d, e)  \\
C$_{64}$(C$_{64}^{+}$), C$_2$   &  9.0 (8.7) & 5.0$\cdot 10^{19}$ (d, e)  \\
C$_{66}$(C$_{66}^{+}$), C$_2$   &  9.4 (9.3) & 5.0$\cdot 10^{19}$ (d, e)  \\
C$_{68}$(C$_{68}^{+}$), C$_2$   &  9.5 (9.5) & 5.0$\cdot 10^{19}$ (d, e)  \\
C$_{70}$(C$_{70}^{+}$), C$_2$   &  9.9 (9.8) & 5.0$\cdot 10^{19}$ (d, e)  \\
All fullerenes, H    &  1.9, 3.3, 4.0 & 5.0$\cdot 10^{19}$ (e,f)  \\
\hline
\end{tabular}
\begin{tablenotes}
\small  
\item References. (a) \cite{murga_orion}; (b) \cite{mic10_hot}; (c) \cite{lebedeva12}; (d) \cite{gluch04}; (e) \cite{matt01}; (f) \cite{bettinger02}. 
\end{tablenotes}
\end{threeparttable}
\end{center}
\end{table*} 

The main parameters required for calculation of $k_{\rm diss}(\rm{index},E)$ are a so-called pre-exponential factor ($k_0$) and activation energy ($E_0$). The pre-exponential factor depends on temperature and the change of entropy of the system during the dissociation ($\Delta S$) (see Eq.~6.80 in \cite{tielens05}). This dependence can be ignored due to its weakness, and $k_0$ can be considered as constant. We adopt the value of $k_0$ estimated by \cite{mic10_hot} both for H and C loss. 

Values of $E_0$ for PAHs are taken to be the same as in our previous work \citep{murga_orion} although another method of calculation of the dissociation rate (the expression in the Arrhenius form) was used. $E_0$ for H loss is consistent with values used in the works of \cite{lepage01, visser07, montillaud13}, where the expression in the inverse Laplace transformation form was used. However, $E_0$ for C loss is quite high in these works. The dissociation rate obtained with such high $E_0$ is too low in comparison with the rates obtained with the expression in the Arrhenius form with $E_0$ adopted in \cite{murga_orion}. We suppose that the parameters from \cite{lepage01, visser07} rather correspond to C loss from centres of PAHs or pure carbon molecules than to C loss from PAHs with hydrogen atoms. Therefore, we use the parameters adopted for the RRKM method in \cite{murga_orion}. The same parameters are used for bPAHs. 

Apart from $k_0$ and $E_0$, the Laplace method requires the set of vibrational frequencies. We adopt frequencies from the NASA Ames PAH database~\citep{boersma14, bauschlicher18}\footnote{\url{https://www.astrochemistry.org/pahdb/}}. Data for some molecules such as coronene (C$_{24}$H$_{12}$) are available not only for the original configuration but also for dehydrogenated and hydrogenated states. In these cases we use vibrational frequencies from the database. However, in most cases there are no frequencies for dPAHs and HPAHs. For HPAHs we adopt the same set of frequencies as for original PAHs with additional number of frequencies near 3~$\mu$m proportional to the additional H atoms. For dPAHs we remove the frequencies around 3~$\mu$m. If PAHs are not fully dehydrogenated, we remove only the number of frequencies proportional to the number of missing H atoms. The database contains the set of frequencies of C$_{66}$, we remove three random frequencies per missing C atom from the set of C$_{62}$ and C$_{64}$. For other bPAHs (C$_{72}$, C$_{74}$, C$_{76}$, C$_{78}$) we rely on the set of frequencies of C$_{78}$H$_{22}$ but we remove frequencies near 3~$\mu$m and three random bands per missing C atom. For fullerenes C$_{58}$, C$_{60}$, C$_{62}$, C$_{64}$, C$_{66}$ we use the set of frequencies for C$_{60}$ regardless of its hydrogenation state, while for C$_{68}$ the set of frequencies of C$_{70}$ is adopted also  regardless the hydrogenation state.

We note that we do not consider H$_2$ detachment for HPAHs as a channel of dissociation, but account for it via barrier-free desorption from HPAHs as done in \cite{andrews16}.

Experiments and theoretical investigations on fullerene dissociation of different sizes were carried out many times~\citep{taylor91, foltin93, matt01, tomita01, gluch04}. Although the discrepancy in estimations of activation energy reaches a factor of 2, the stability of fullerenes relative to PAHs is undisputed. $E_0$ of fullerenes varies from $\approx7$ to 10~eV depending on a cage and estimation method while $E_0$ for PAHs is $\approx4-5$~eV. We adopt $E_0$ for fullerenes from \cite{gluch04}, while $k_0$ is taken from \cite{matt01}.

Besides C loss, fullerenes can lose H atoms if they are hydrogenated. The H detachment from fulleranes was theoretically investigated. In particular, \cite{bettinger02} studied thermally induced detachment of H$_2$. They considered four reactions: C$_{60}$H$_2\to{\rm C}_{60}$H+H,  C$_{60}$H$_2+{\rm H}\to{\rm C}_{60}$H$+$H$_{2}$, C$_{60}{\rm H}\to{\rm C}_{60}$+H,  C$_{60}$H$+{\rm H}\to{\rm C}_{60}$H$_2$. They estimated the rate of H$_2$ loss, when all these reactions take place at once, and obtained the parameters for the Arrhenius law for the mix of these reactions. We take their activation energies for individual dissociation reactions (three values presented in Table~\ref{par} separated by commas) and adopt the pre-exponential factor to be the same as for C loss. \cite{cataldo09b} investigated the influence of UV radiation on fulleranes and measured rate constants of dehydrogenation. They showed that the rates increase as hydrogenation level increases. These rate constants cannot be directly incorporated into our model, therefore we use parameters obtained for thermolysis which dependence on the hydrogenation level has not been investigated. Due to the lack of available data we use the set of frequencies of fullerenes for calculations of H loss rate of fulleranes in spite of differences in their IR spectra~\citep{cataldo09b, iglesias12, cataldo14}.

\subsection{Folding}

Bending of dPAHs and consequent folding into fullerenes are triggered by dissociation and therefore treated in a similar way. This process has been investigated both experimentally and theoretically~\citep{chuvilin10, lebedeva12, zhen14_ful, pietrucci14}. However, both qualitative and quantitative descriptions are still difficult. We drew an evolutionary scheme that is based on the results of recent experiments and modelling with tools of the molecular dynamics which are available to date.

\cite{zhen14_ful} showed that the PAH bending and folding into fullerenes is quite efficient, but not the only channel of PAH dissociation. Based on this experiment it was estimated that the maximum fraction of PAHs, which turn into fullerenes upon irradiation, can reach about 20\%. Another fraction of PAHs likely keep loosing carbon atoms turning into smaller PAHs or carbon chains. This fact reveals that the activation energy of the bending must be comparable with the activation energy of C loss. \cite{berne15} assumed that the planar C$_{66}$ folds to the cage C$_{66}$. They emphasised that this channel dominates over other channels, and most of the planar C$_{66}$ turn into cages. Subsequently losing C$_2$ the cage C$_{66}$ converts to smaller cages C$_{64}$, C$_{62}$, C$_{60}$, C$_{58}$. The cage C$_{60}$ exists longer due to its relatively high stability. However, according to \cite{gluch04} the activation energy of all these cages are high (8--10~eV). We approximately estimated the rates of conversion of the cages into smaller ones under the ISM conditions and concluded that this conversion is very slow, i.e. evolution of the cages will nearly stop at the first cage (C$_{66}$). The results presented in the work of \cite{zhen14_ful}, on the other hand, indicate that C$_{60}$ must be the dominant among other C$_{x}$. So we suppose that the conversion from C$_{66}$ to C$_{60}$ should go with higher rates than those predicted by the scenario of \cite{berne15}.  

We propose the scheme that differs from \cite{berne15} in some points. It includes the following steps:
\begin{itemize}
\item{dehydrogenation of PAHs;}
\item{bending and C$_{2}$ loss from C$_{66}$ and C$_{78}$, which ends up with folding into the cages C$_{60}$ and C$_{70}$, correspondingly. Herewith we emphasise that bending competes with fragmentation, i.e. not all dPAHs will fold into cages. The probability of the channel is determined by relation between its rate and rates of other competing processes;}
\item{shrinking of C$_{70}$ to smaller cages, including C$_{60}$, due to C$_2$ loss. However, energies required for this process are quite high.}
\end{itemize}

The rate which determines the probability of the bending channel ($I_{ijk}$) can be calculated using Eq.~\ref{r_destr}. Following  \cite{berne15} we adopt the same parameters $k_0$ and $E_0$ from the work of \cite{lebedeva12}, who investigated folding of dPAH C$_{96}$ and C$_{384}$. The activation energies and pre-exponential factors of these 2 dPAHs  are very close to each other in spite of the difference in the number of C atoms. So we suppose that the parameters for C$_{66}$ and C$_{78}$ are not far from the parameters for C$_{96}$, consequently, we adopt $E_0=4.0$~eV and $k_0=1.0\cdot 10^{15}$~s$^{-1}$ for bending of dPAHs. 

Once dPAHs entered the bending channel, their following dissociation is determined by coefficients $B_{k}$. We assume that bPAHs have the same stability as PAHs unlike \cite{berne15} who adopted fullerene stability for bPAHs. Thus, we adopt C$_2$ loss from bPAHs C$_{66}$ to C$_{60}$ (or C$_{78}$ to C$_{70}$) proceeds with the same activation energy as for other PAHs, which we adopt to be equalled to $\approx4.6$~eV. As soon as the cages C$_{60}$ and C$_{70}$ are reached, the molecules acquire the fullerene stability which is much stronger than stability of PAHs. We admit the $E_0$ of dehydrogenated PAHs is likely higher than for PAHs and closer to $E_0$ of fullerenes, but the exact value is undetermined, so we consider our rates of dissociation of bPAHs as an upper limit. The coefficient $B_{k}$ corresponding to the last step of a bPAH evolution that leads to folding to fullerenes equals to the coefficient $F_{klp}$.

\cite{lebedeva12} demonstrated that nickel can be a catalyst for the folding of dPAHs to fullerenes. In this case the activation energy decreases from 4~eV to less than 2~eV. It is likely that some other atoms abundant in the ISM (e.g. nitrogen) may lead to the same results, and the folding can be much faster in the ISM than it is assumed in this work. More experimental and theoretical investigations are needed. 

\subsection{Hydrogenation}
\label{hydro}

Hydrogenation of PAHs is calculated analogously to our previous work~\citep{murga_acet}. Here we describe the hydrogenation of fullerenes, i.e. formation of fulleranes. 

Fullerenes are considered as a lightweight storage of hydrogen inside their cages, therefore, the interaction between fullerenes and hydrogen is studied extensively. The extraction of hydrogen solely from fulleranes seems most plausibly as C-H bonds are weaker than C-C bonds. Thus, increasing the fullerene temperature or energy up to a certain level leads to evaporation of hydrogen atoms without destruction of the carbon skeleton.  

Experiments indicate that implantation of hydrogen onto fullerenes can proceed efficiently~\citep{howard93, petrie95, thrower19}. During the interaction double C-C bonds are broken and new C-H bonds are formed. Most likely hydrogen atoms are attached outside of the fullerene cage, the addition occurs with no barrier while the penetration of hydrogen inside the cage has a high barrier~\citep{vehlilainen11}. Adsorption of the H$_2$ molecule is also possible if its energy is high enough to ensure its dissociation during contact and subsequent H atom adsorption ~\citep{vehlilainen11}.  Collisions of H$_2$ with fullerenes with low energy would rather lead to its physisorption~\citep{kaiser13}. Therefore, we do not consider the attachment of the H$_2$ molecule as these molecules are present in the ISM at temperature lower than it is required for attachment to fullerenes. 
 
\cite{petrie95} experimentally estimated the rates of H addition to ionised fullerenes. The lower limit for a singly ionised C$_{60}$ is around 10$^{-10}$~cm$^3$~molecule~s$^{-1}$. The experiments revealed the efficient hydrogen attachment to neutral C$_{60}$ as well, although there are no estimations of the rates~\citep{howard93, thrower19}. Accordingly, we adopt the same rate for C$_{60}$ and all other fullerenes as for C$_{60}^+$. In mathematical terms, $H^{lp}$ for all $l,p$ and all ionisation states equals 10$^{-10}$~cm$^3$~molecule~s$^{-1}$.
 
It should be noted that saturation of fullerenes with hydrogen lead to the overall reduction of stability of their carbon skeleton~\citep{luzan11}, however, in this work, we assume that fulleranes may lose only their hydrogen atoms. 

\subsection{Properties of PAHs, bowl-shaped PAHs, and fullerenes}

Our model requires knowledge about physical properties of PAHs, bPAHs and fullerenes such as absorption cross-section, heat capacity, photo-ionisation yield, rates of electron recombination. In this section, we summarise these properties.

\subsubsection{Absorption cross-sections}

The absorption cross-sections for PAHs are taken from \cite{DL07}, while the cross-sections for dPAHs and HPAHs were obtained in our previous work by scaling strength of the IR bands related with vibrations of C-H bonds proportionally to the number of H atoms~\citep{murga_orion}. 

The absorption cross-sections for fullerenes were compiled by combining the results of several works devoted to different spectral ranges. The UV cross-section of C$_{60}$ in the wavelength range from 300 to 2500~\AA{} was presented in \cite{berkowitz99}, while the range from 2500 to 6000~\AA{} is presented in \cite{coheur96}. We use these cross-sections for cages C$_{66}$, C$_{64}$, C$_{62}$, C$_{58}$ as well but with scaling proportional to the number of C atoms according to \cite{berne15, joblin92}. The absorption cross-section of C$_{70}$ is available only for the wavelength range from 2500 to 6000~\AA{} in \cite{coheur96}. For the shorter wavelengths we use data for C$_{60}$, scaling them proportionally to the number of C atoms. We scale the cross-section of C$_{70}$ for the cage C$_{68}$. 

We calculate the IR cross-sections based on the parameters for vibrational frequencies presented in the NASA Ames PAH database. In the database, the parameters are given only for C$_{60}$ and C$_{70}$, so we scale the cross-sections for all the other cages as we do for the UV cross-sections.

\subsubsection{Photoionisation}

The photoionisation yield for PAHs including dPAHs, bPAHs, and HPAHs is calculated according to \cite{wd01_charging}. The photoionisation yield for C$_{60}$ is taken from \cite{berkowitz99}. Due to high ionisation potential of C$_{60}^{+}$ (about 18~eV; \cite{pogulay04}), the second ionisation is unlikely under conditions of PDRs, where photon energy is limited by 13.6~eV, therefore, we consider only the first ionisation (C$_{60}$ to C$_{60}^+$). Besides the direct ionisation, thermionic emission is possible, when the internal energy of fullerenes is high enough ($\approx30$~eV), but we neglect this process due to its minor role as demonstrated by \cite{berne15}. The ionisation potential of C$_{70}$ and other fullerenes is very close to the ionisation potential of C$_{60}$~\citep{boltalina00}, so we adopt the photoionisation yield of C$_{60}$ for all fullerenes.

\subsubsection{Interaction with electrons}

\cite{wd01_charging} showed that the sticking coefficient for C$_{60}$ is lower than for PAHs at least at low temperatures. \cite{viggiano10} measured and modelled the rate coefficient for the electron attachment to neutral C$_{60}$, and we adopt this coefficient. There are no estimates of the attachment cross section for negatively charged fullerenes, and electron recombination rate for positively charged fullerenes are uncertain, therefore we adopt the same rate coefficients for positively and negatively charged fullerene ions as for neutral C$_{60}$. 


We take photo-detachment cross-section for all fullerenes from \cite{amusia98}. The electron affinity of C$_{60}$ is adopted to be 2.7~eV~\citep{boltalina97}. This value for C$_{70}$ and other fullerenes is very close to C$_{60}$, therefore, we adopt this value for them, too. 

\subsubsection{Heat capacity}

To calculate heat capacity $\mathcal{C}$,\,erg\,K$^{-1}$ of PAHs we rely on the work of \cite{DL01_stoch}. We fit their PAH heat capacities per C atom by the temperature dependent function suggested in \cite{pavyar12}:
\begin{equation}
\frac{\mathcal{C}}{N_{\rm C}k_{\rm B}} = \frac{3}{1+\left(\frac{T_{\star}}{T_{\rm d}}\right)^2},
\end{equation}
where T$_{\rm d}$ is a molecule temperature, $k_{\rm B}$ is the Boltzmann constant, and $T_{\star}$ is a parameter.

Heat capacity of C$_{60}$ was measured by \cite{diky01} in the temperature range from 100 to  
640~K, while theoretical estimations for C$_{60}$ and C$_{70}$ are available for the whole range up to 1000~K, which are consistent with the available experimental data~\citep{diky00}. We fit the combination of theoretical and experimental data (or only theoretical data in case of C$_{70}$) by the function consisting of two terms, which are analogous to the function used for PAHs: 
\begin{equation}
\frac{\mathcal{C}}{N_{\rm C}k_{\rm B}} = \frac{A}{1+\left(\frac{T_{\star}^{(1)}}{T_{\rm d}}\right)^2} + \frac{3}{1+\left(\frac{T_{\star}^{(2)}}{T_{\rm d}}\right)^2},
\end{equation}
where $A=0.11$, $T_{\star}^{(1)}=17.2$~K and $T_{\star}^{(2)}=470$~K are parameters, which are obtained from the fitting procedure. The data for C$_{60}$ and C$_{70}$ are very close to each other, therefore, we use this function for all the considered fullerenes. 

\subsection{Model parameters}

We perform the modelling in two ways. First, we consider two representative environments, the Orion Bar and NGC~7023 PDRs. Second, we model over a grid of parameters: intensity of radiation field (we designate it by $U$, which represents a scale factor of the radiation field suggested in \citealt{mmp83}), the gas temperature $T_{\rm gas}$, atomic hydrogen number density $n({\rm H})$, and electron number density $n_{\rm e}$. We perform the modelling separately with inclusion of hydrogenation (this case is designated `Hydr.$+$') and without it (`Hydr.$-$') as this process substantially changes the results although its efficiency and rate are uncertain. 

We model the Orion Bar PDR for $A_{\rm V}$ from 0.1 (the ionisation front, IF) to 10 (interior of the molecular cloud). We adopt the parameters for the Orion Bar from \cite{goicoechea15}, and parameters for NGC~7023 were taken from \cite{berne15}. Radial dependence of these parameters on the distance from the ionising star can be found in the cited papers. 

\begin{figure}
\includegraphics[width=0.49\textwidth]{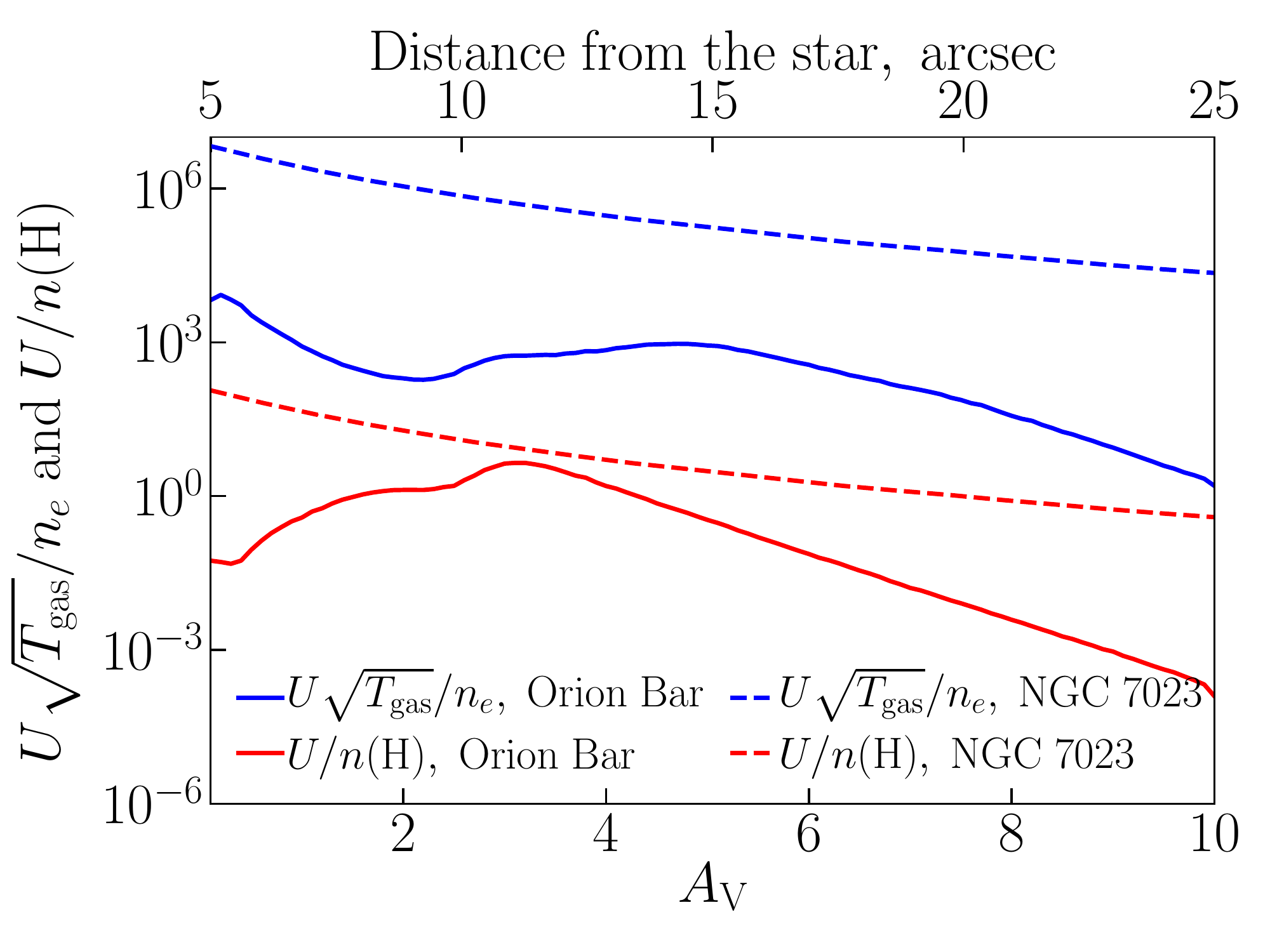}
\caption{The parameters $U \sqrt{T_{\rm gas}} / n_{\rm e}$ (blue lines) and $U /n({\rm H})$ (red lines) vs. extinction $A_{\rm V}$ for the Orion Bar and vs. distance from the star for NGC~7023. Solid lines are for the Orion Bar, dashed lines are for NGC~7023.}
\label{fig: gtn}
\end{figure}

We use two parameters, the combination of the radiation field intensity, the gas temperature and the electron density, $U \sqrt{T_{\rm gas}}/n_e$, and the ratio between $U$ and $n({\rm H})$. The first parameter determines the ionisation balance of the molecules as it connects the three main factors influencing the charge. The larger this parameter is, the farther the charge distribution function shifts towards positive ions. The second parameter roughly characterises the hydrogenation level of PAHs and fullerenes. The higher the ratio, the lower the hydrogenation level of the molecules. In other words, if the ratio is lower than some specific value (which depends on the molecule type), the hydrogenation can compensate the H loss. Radial dependencies of these parameters on the distance from the ionising star in the Orion Bar and NGC~7023 are presented in Fig.~\ref{fig: gtn}. 
 
Our grid of parameters includes the ranges of $U$ from 0.1 to 10$^{6}$, $T_{\rm gas}$ from 10 to 10$^4$~K, $n({\rm H})$ from 10$^{-2}$ to 10$^5$~cm$^{-3}$, and $n_{\rm e}$ from 10$^{-3}$ to 10$^{2}$~cm$^{-3}$ (we limit $n_{\rm e}$ so that it does not exceed $n({\rm H})$). Our grid presumably covers the significant fraction of possible combinations of parameters in the ISM including PDRs, WIM, WNM, CNM, PNe. We perform the calculations throughout the grid to find the combination of parameters favourable for the most efficient fullerene formation. 

The initial PAH size distribution is taken from \cite{wd01} for $R_{\rm V}=3.1$ and the maximum carbon abundance. PAHs lock about 7\% of all carbon atoms. 

\section{Model results}

In this section, we consider the results of our modelling. We consider separately the results of the modelling in the Orion Bar, NGC~7023 and on the grid of parameters. We pay more attention to the Orion Bar and consider only specific interesting points for NGC~7023 and parameter grid case. 

\subsection{The Orion Bar}

\begin{figure}
\includegraphics[width=0.49\textwidth]{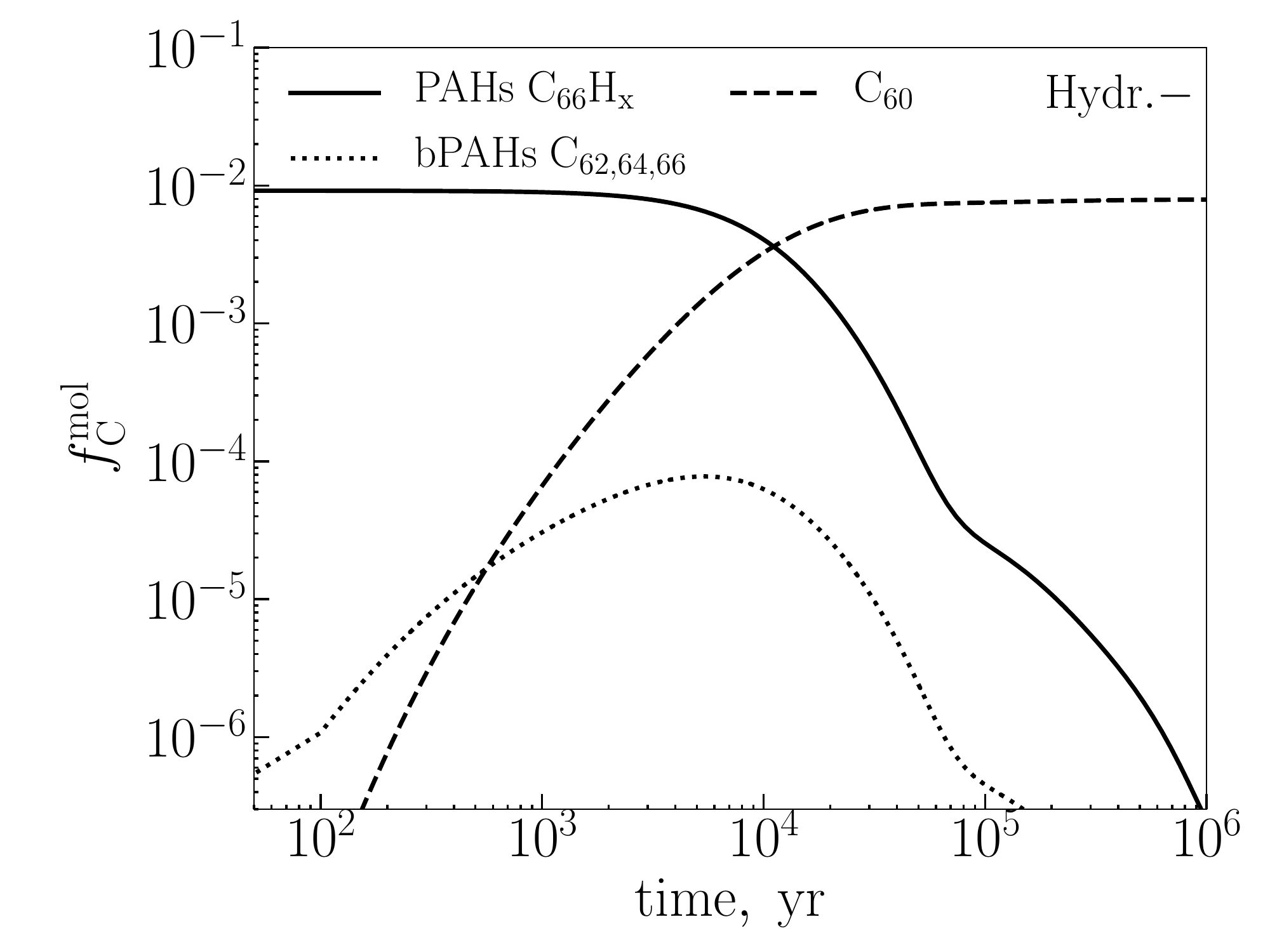}
\caption{Dependence of the number density of fullerene C$_{60}$, sum of bPAHs C$_{62}$, C$_{64}$, C$_{66}$, and PAH C$_{66}$H$_x$ with $x \geq 0$ on time. The profiles correspond to $A_{\rm V}=0.9$. The hydrogenation is not included.}
\label{fig: nful_time}
\end{figure}

Based on our modelling we calculate the fraction of carbon locked in different types of molecules ($f_{\rm C}^{\rm mol}$) relative to the total carbon abundance (C, C$^{+}$, CO and carbonaceous dust including PAHs). In Fig.~\ref{fig: nful_time} we show evolution of $f_{\rm C}$ in fullerene C$_{60}$ and in its precursors: C$_{66}$H$_{20}$, bPAHs (sum of $f_{\rm C}$ of C$_{62}$, C$_{64}$, C$_{66}$). Profiles on the figure reflect the sequence of processes, which were incorporated in our model. Firstly, the original PAH C$_{66}$H$_{20}$ loses its hydrogen atoms due to photo-dissociation. Being dehydrogenated, it may start folding into a cage, and at this stage we call it bPAH. In its turn, this stage consists of three consecutive substages (C$_{66}$, C$_{64}$, C$_{62}$), which the molecule passes losing carbon atoms. After these stages, the fullerene C$_{60}$ appears, and it dominates over the original PAH after 10$^{4}$~yr.

\begin{figure*}
\includegraphics[width=0.9\textwidth]{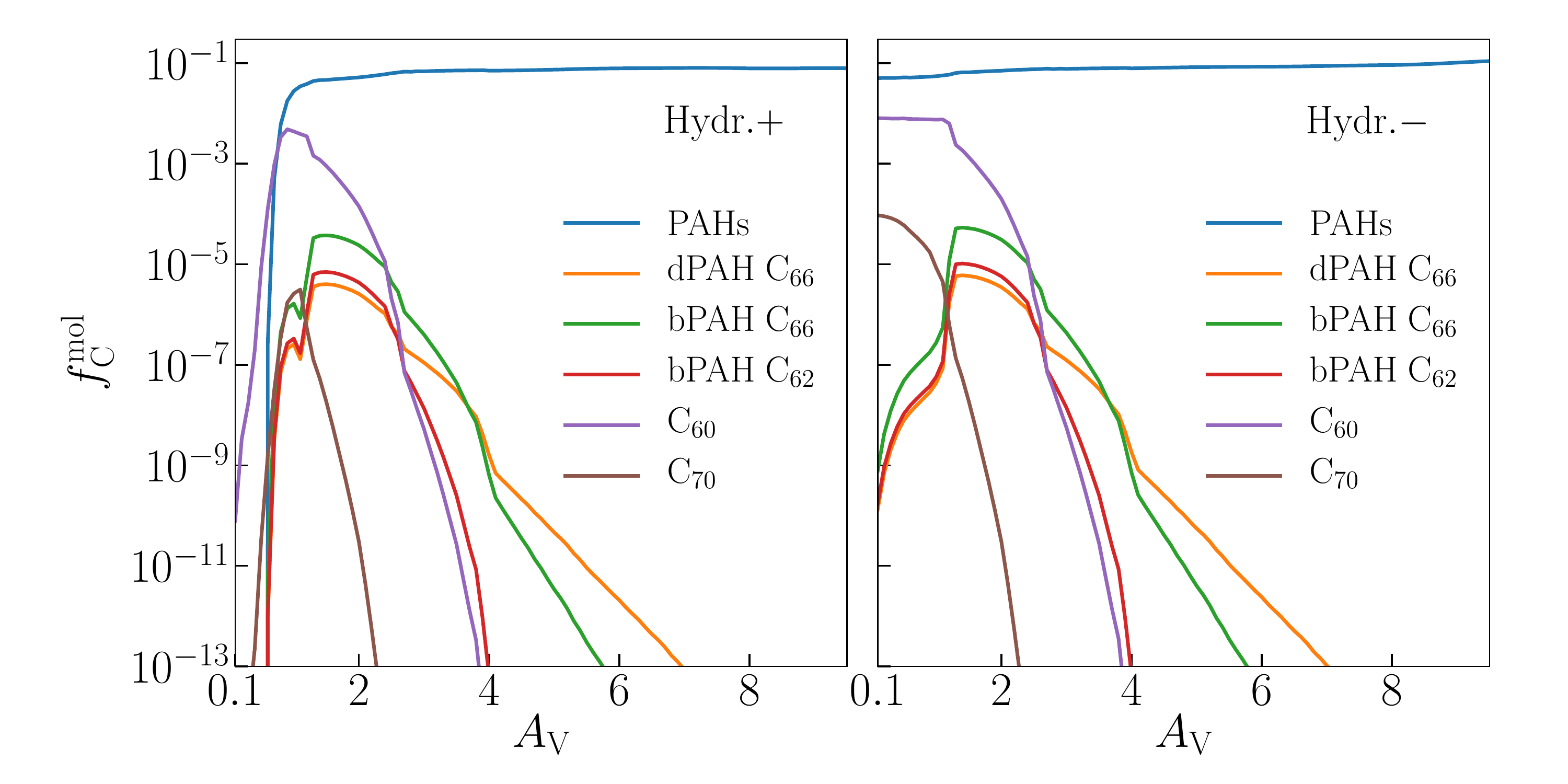}
\caption{The fractions of carbon locked in PAHs (i.e. HPAHs + normal PAHs + dPAHs), dPAH C$_{66}$, bPAHs C$_{66}$ and C$_{62}$, and fullerenes C$_{60}$ and C$_{70}$ versus
$A_{\rm V}$ are presented. The hydrogenation is included into the modelling on the left panel and is not included on the right panel}
\label{fig: nall_laplace}
\end{figure*}

In Fig.~\ref{fig: nall_laplace} we present the fractions of carbon locked in all PAHs, dPAH C$_{66}$, bPAHs C$_{66}$ and C$_{62}$, fullerenes C$_{60}$ and C$_{70}$ versus $A_{\rm V}$ in the Orion Bar at 10$^5$~yr, which is an approximate age of the PDR~\citep{salgado16}. The calculations start at $A_{\rm V}=0.1$, which corresponds to the position of the IF. We show the results of the modelling with (left) and without (right) accounting for the hydrogenation.

On the left panel the deep trough in $f_{\rm C}$ for all molecules is seen at $A_{\rm V}<0.5$. $f_{\rm C}^{{\rm C}_{60}}$ grows faster than $f_{\rm C}$ in other molecules. The maximum $f_{\rm C}^{{\rm C}_{60}}$ about $4\cdot 10^{-3}$ is reached at $A_{\rm V}\approx 0.9$. It starts falling steeply after $A_{\rm V}\approx1.5$. $f_{\rm C}^{{\rm C}_{70}}$  also has the maximum at $A_{\rm V}\approx 0.9$ but the value is only about $10^{-6}$. $f_{\rm C}^{{\rm C}_{60}}$ is superior to $f_{\rm C}^{\rm PAHs}$ in the range of $A_{\rm V}\lesssim 0.5$ and to $f_{\rm C}$ in dPAH C$_{66}$ and in bPAHs at $A_{\rm V}\lesssim 2.5$. Also, $f_{\rm C}$ in dPAH C$_{66}$ is lower than $f_{\rm C}$ in bPAH C$_{66}$ at $A_{\rm V}\lesssim 4$. This relation between the fractions follows from the relation between the rates of their formation. By the age of 10$^5$~yr, nearly all PAHs of this size are dehydrogenated. The conversion of dPAHs into bPAHs is also very efficient, and its rate is larger than the rate of fragmentation, which does not lead to bending. Thus, the significant fraction of dPAHs transform to bPAHs, which in its turn quickly fold up into fullerenes under harsh radiation field. At $A_{\rm V}\gtrsim 2.5$  $f_{\rm C}^{{\rm C}_{60}}$ becomes lower than $f_{\rm C}^{{\rm dPAH}}$ and $f_{\rm C}^{{\rm bPAH}}$. At $A_{\rm V}\gtrsim 2.5$ $f_{\rm C}^{{\rm PAHs}}\approx7\cdot 10^{-2}$, which corresponds to the initial value obtained from the WD01 distribution. 

Let us now take a look at the right panel of Fig.~\ref{fig: nall_laplace}, where hydrogenation is not included. A value of $f_{\rm C}^{{\rm C}_{60}}$ reaches its maximum ($\approx7\cdot10^{-3}$) at $A_{\rm V}=0.1$, remains at about the same level up to $A_{\rm V}\approx1.5$, and then sharply drops. A value of $f_{\rm C}^{{\rm C}_{70}}$ also has its maximum ($\approx8\cdot 10^{-5}$) at $A_{\rm V}=0.1$, but starts falling right after the IF. A value of $f_{\rm C}^{\rm PAHs}$ is $\approx 4\cdot 10^{-3}$ at $A_{\rm V}=0.1$, thus, C$_{60}$ abundance is about one fifth of the PAH abundance, and up to 0.7\% of carbon is locked in fullerenes at $A_{\rm V}=0.1$ (sum of C$_{60}$ and C$_{70}$). We emphasise that the minimum $f_{\rm C}$ in PAHs corresponds to the maximum $f_{\rm C}$ in C$_{60}$ and C$_{70}$. At $A_{\rm V}\gtrsim 1$ the profiles on the left and right panels of Fig.~\ref{fig: nall_laplace} coincide as the hydrogenation becomes inefficient. 

\begin{figure}
\includegraphics[width=0.49\textwidth]{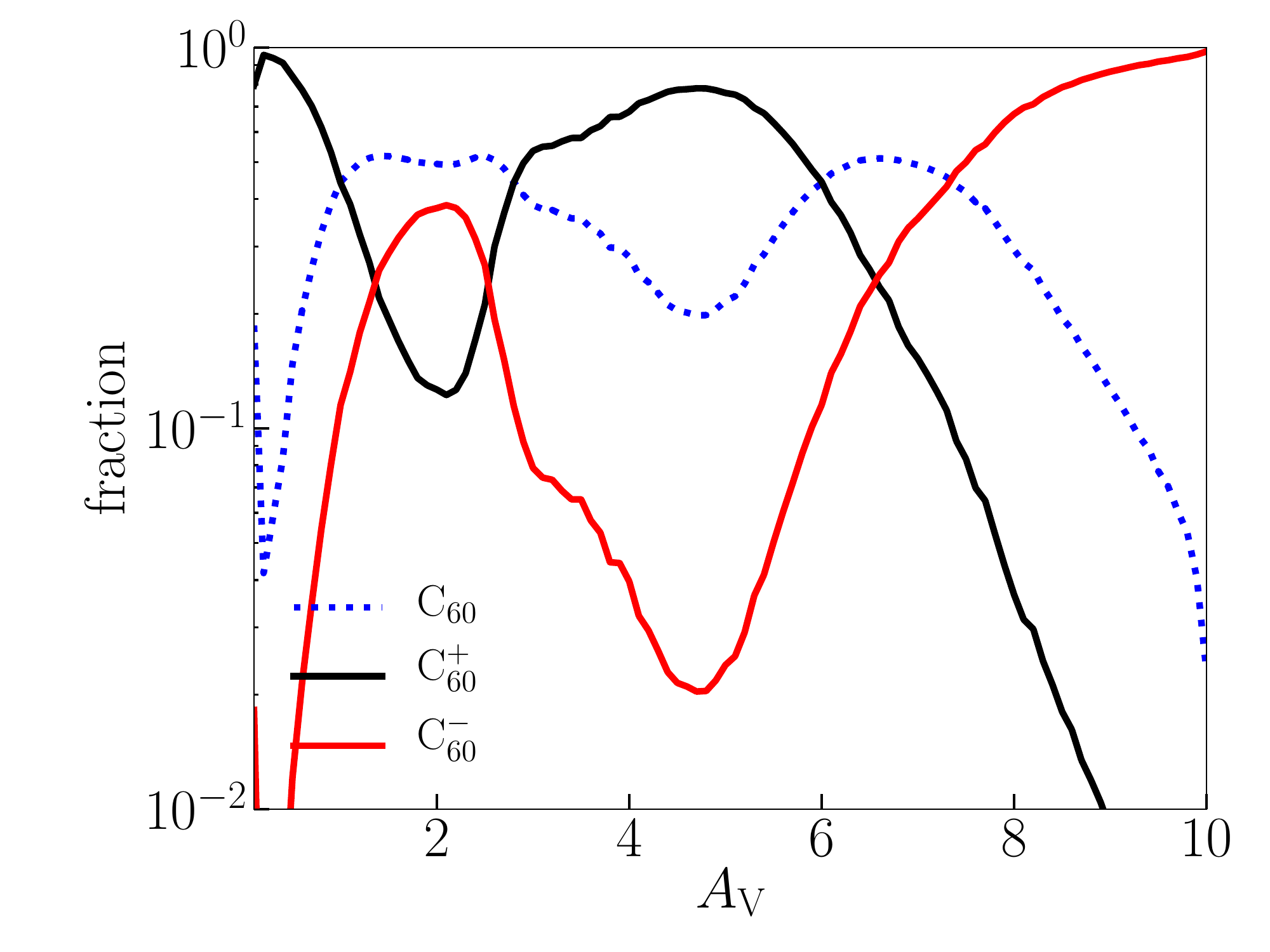}
\caption{Fraction of C$_{60}$, C$_{60}^{+}$ and C$_{60}^{-}$ vs. $A_{\rm V}$ in the Orion Bar.}
\label{fig: nful_ion}
\end{figure}

In Fig.~\ref{fig: nful_ion} we show the ionisation state of fullerenes throughout the Orion Bar. At $A_{\rm V}\lesssim 1.0$ C$_{60}^+$ dominate due to high radiation field intensity and efficient photo-ionisation. Within the range of $1<A_{\rm V}<3$ C$_{60}$ dominate, though the fraction of C$_{60}^-$ is also noticeable. These high abundances of C$_{60}$ and C$_{60}^{-}$ relative to C$_{60}^{+}$ reflect the profile of the parameter $U \sqrt{T_{\rm gas}} / n_e$, where it has a minimum, i.e. charging by electrons compensates photo-ionisation. Deeper in the cloud, the fraction of C$_{60}^{+}$ again rises, but at $A_{\rm V} \geqslant 6$ ions C$_{60}^{-}$ dominate.  

\begin{figure}
\includegraphics[width=0.49\textwidth]{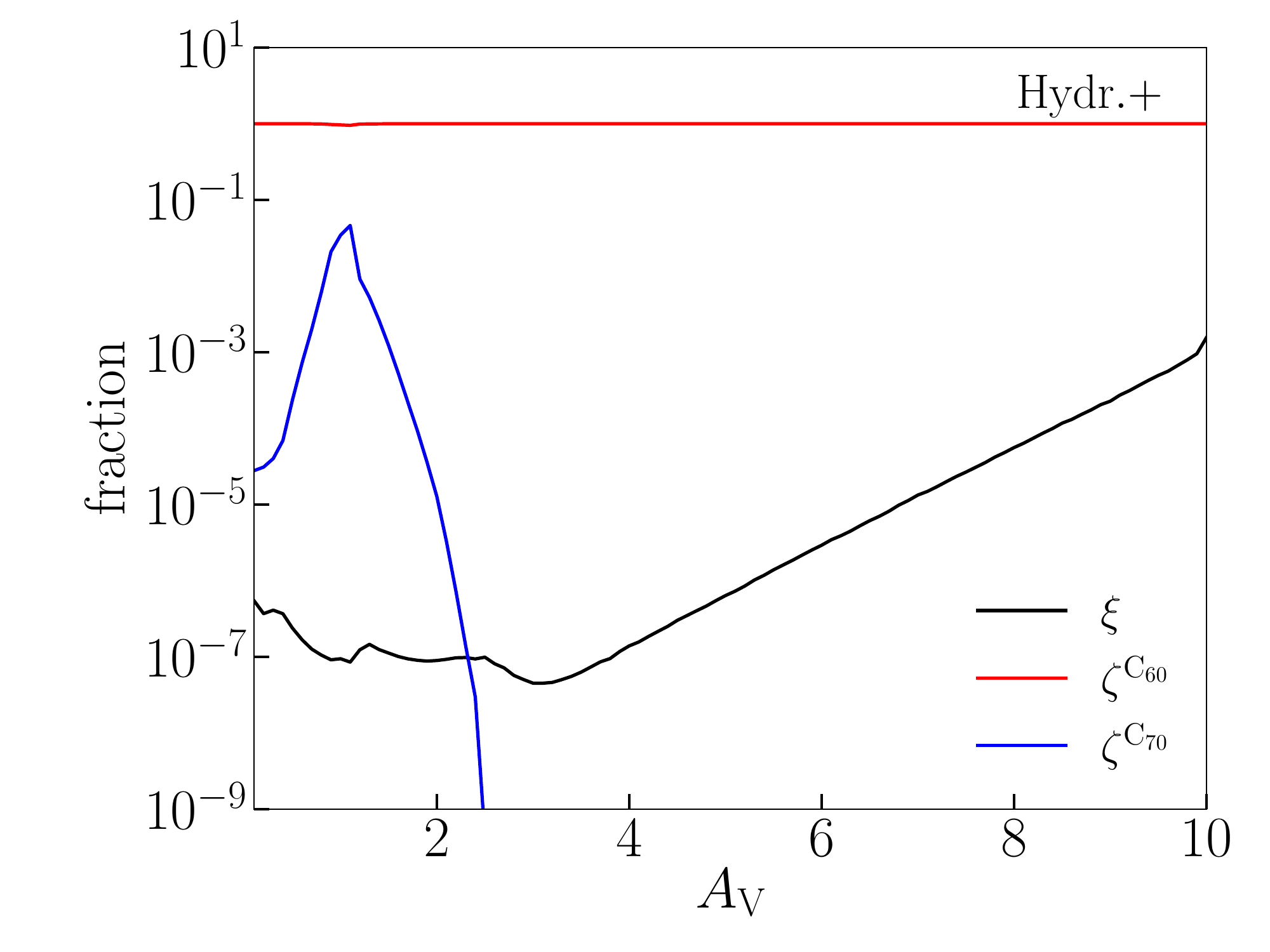}
\caption{The fractions of C$_{60}$ ($\zeta^{{\rm C}_{60}}$) and C$_{70}$ ($\zeta^{{\rm C}_{70}}$) among fullerenes of all sizes and hydrogenation states, and the fraction of fulleranes C$_{60}$H$_x$ ($\xi$) among C$_{60}$ of all hydrogenation states vs. $A_{\rm V}$. $\zeta = N_{21}^{\rm ful}/\sum\limits_{lp}N_{lp}^{\rm ful}$, $\xi = \sum_{p=2..10}N_{2p}^{\rm ful}/\sum_{p=1..10}N_{2p}^{\rm ful}$. The hydrogenation is included.}
\label{fig: nful_hydr}
\end{figure}

In Fig.~\ref{fig: nful_hydr} the fractions of C$_{60}$ and C$_{70}$ among fullerenes of all sizes and hydrogenation states (designated as $\zeta^{{\rm C}_{60}}$ and $\zeta^{{\rm C}_{70}}$), and also the fraction of fulleranes C$_{60}$H$_x$ among fullerenes C$_{60}$ of all hydrogenation states (designated as $\xi$) are demonstrated. It is clear that the dominant fullerene is C$_{60}$, while the fraction of C$_{70}$ is much lower. The fraction of C$_{70}$ reaches the maximum value at $A_{\rm V}\approx 1$. The fraction of fulleranes is even lower than the fraction of C$_{70}$. It increases at $A_{\rm V}>5$ but still does not exceed $10^{-3}$. 

\begin{figure}
\includegraphics[width=0.49\textwidth]{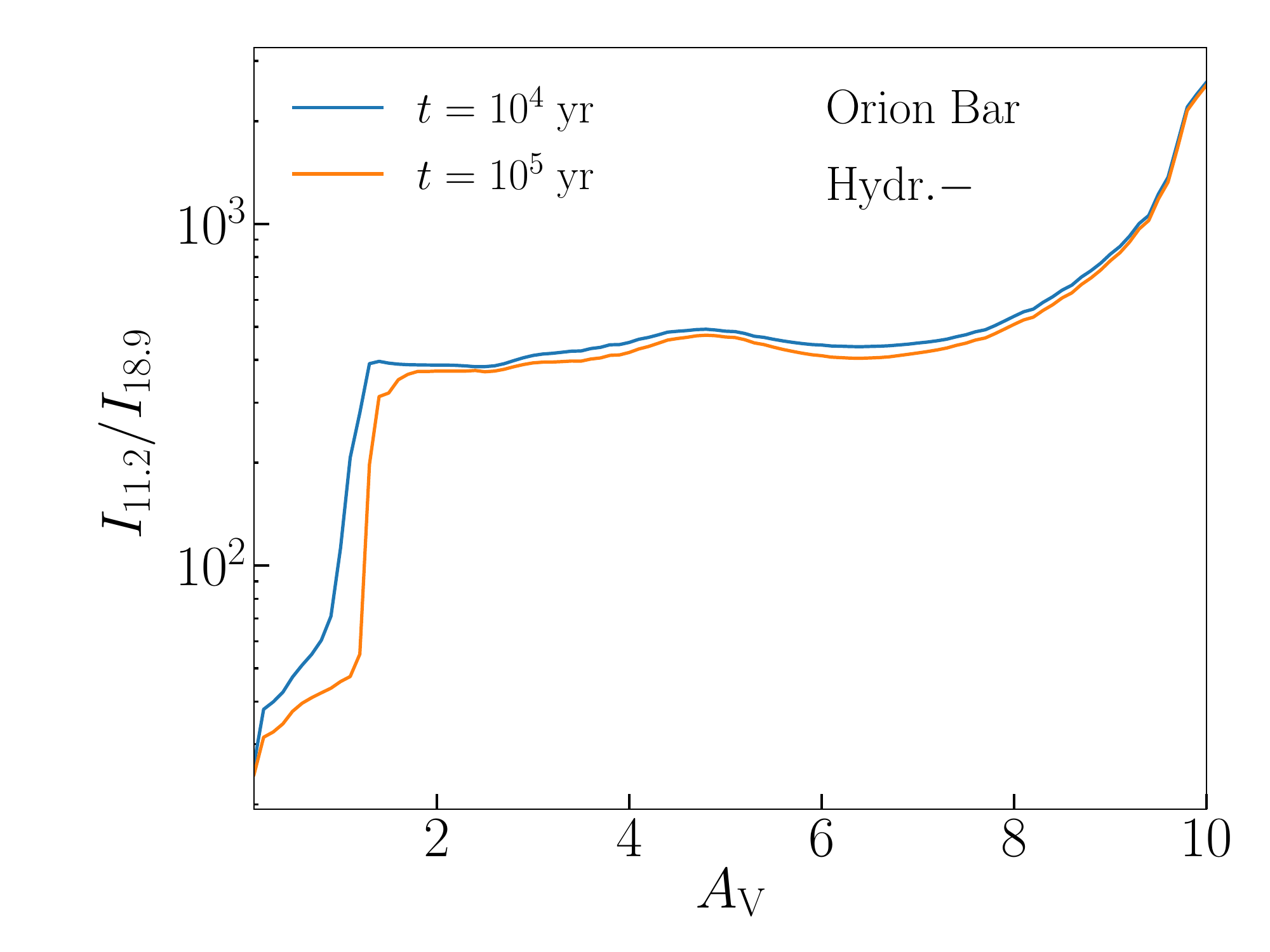}
\caption{The dependence of the ratio $I_{11.2}/I_{18.9}$ on $A_{\rm V}$ in the Orion Bar at $t=10^4$ and $10^5$~yr. The hydrogenation is not included.}
\label{fig: ratio_av_orion_laplace}
\end{figure}

Based on the obtained abundances of carbonaceous molecules we calculated IR spectra of the Orion Bar with the {\tt DustEm} emission model~\citep{dustem}. We calculated synthetic intensities of the bands at 11.2 and 18.9~$\mu$m. Regarding the 11.2~$\mu$m band, we applied the same procedure of subtracting of global continuum and underlying plateau, which we used in the work of \cite{murga_orion}. To extract the 18.9~$\mu$m band, we subtracted the global continuum that was obtained by spline fitting the points at 10, 15, and 20~$\mu$m. It is assumed that the 18.9~$\mu$m band extends from 18.5 to 19.3~$\mu$m. We illustrate the ratio between intensities of 11.2 and 18.9~$\mu$m bands along the Orion Bar in Fig.~\ref{fig: ratio_av_orion_laplace} for 10$^{4}$ and 10$^5$~yr. The ratio has a minimum equal to $\approx 25$ at $A_{\rm V}=0.1$, then steeply increases until $A_{\rm V}\approx 1.5$, where the ratio is around 400, and slowly increases farther up to the value of few thousands. The ratio at 10$^5$~yr is slightly lower than the ratio at 10$^4$~yr, indicating gradual transformation of PAHs to fullerenes.  

\subsection{NGC 7023}

\begin{figure*}
\includegraphics[width=0.45\textwidth]{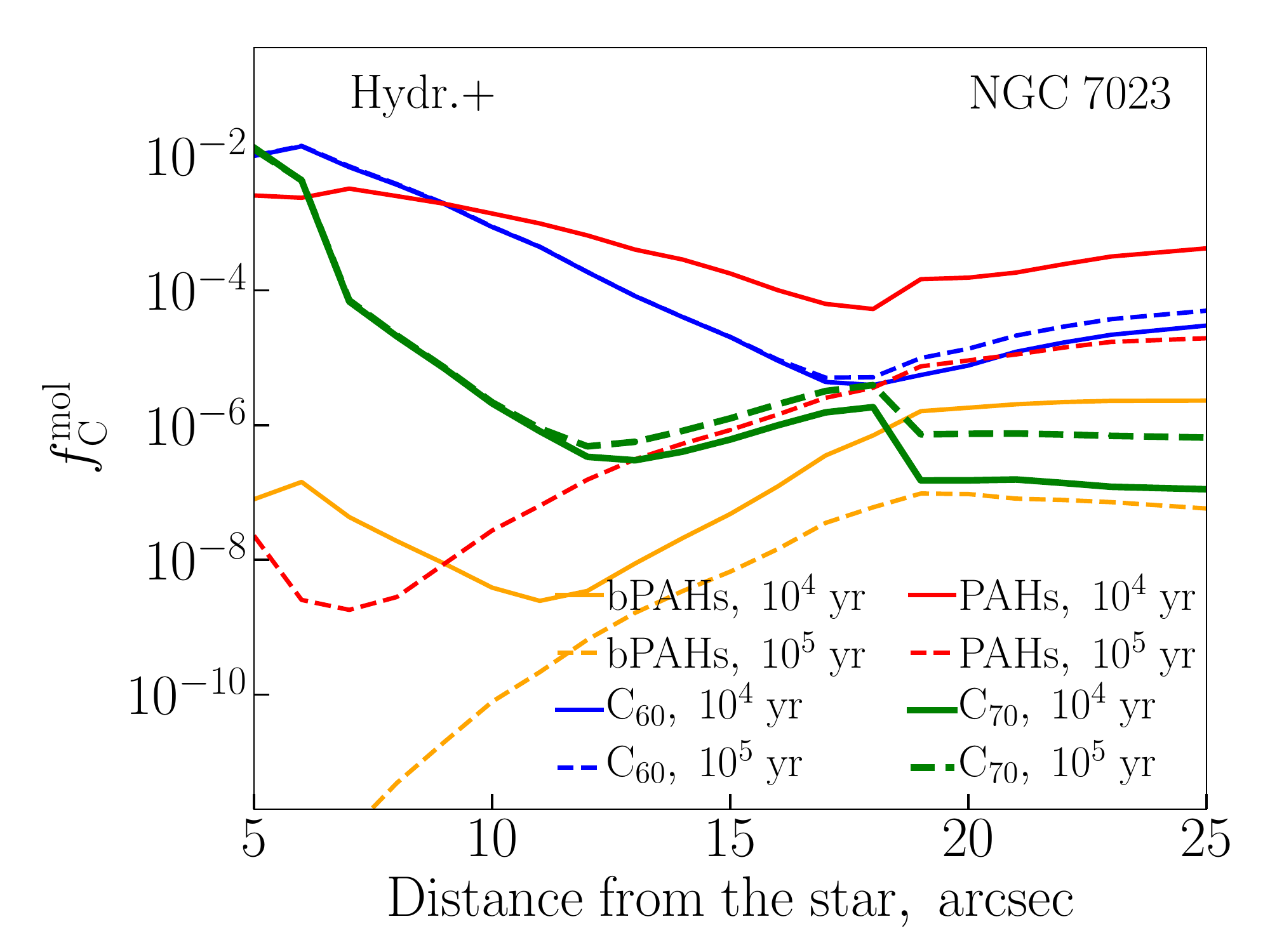}
\includegraphics[width=0.45\textwidth]{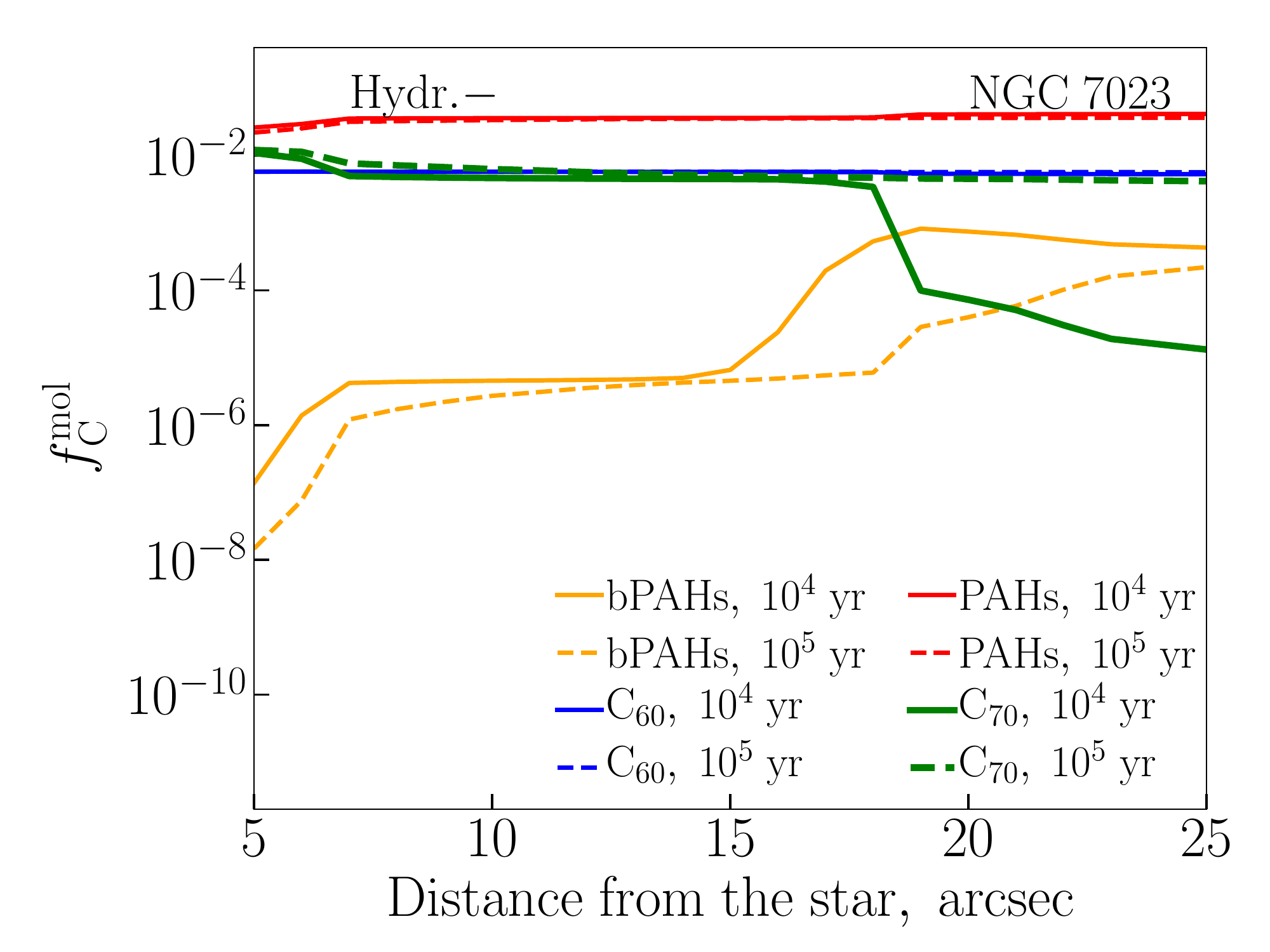}
\caption{The fraction of C locked in C$_{60}$, C$_{70}$, bPAHs, and PAHs in NGC~7023 vs. distance from the star. The hydrogenation is included on the left panel and is not included on the right panel. The solid lines correspond to the age of 10$^4$~yr, and the dashed lines correspond to 10$^{5}$~yr.}
\label{fig: fc_ngc7023}
\end{figure*}

The NGC~7023 case allows extending the range of considered conditions for fullerene formation as this region is characterised by higher values of $U \sqrt{T_{\rm gas}} / n_{\rm e}$ and $U /n({\rm H})$. In Fig.~\ref{fig: fc_ngc7023} the dependence of $f_{\rm C}$ locked in PAHs and fullerenes on $A_{\rm V}$ is presented. We show the results of modelling with included hydrogenation on the left panel and without hydrogenation on the right panel. The age of NGC~7023 is estimated to be between 10$^4$ and 10$^5$~yr~\citep{berne15}, therefore, we present the results for both ages. The radiation field is quite intense in this object and at the brightest locations is comparable with the Orion Bar. On the right panel $f_{\rm C}^{\rm PAHs}$ fluctuates around $7\cdot10^{-2}$, having minimum at $5\arcsec$ and maximum at $25\arcsec$. A value of $f_{\rm C}^{{\rm C}_{60}}$ at $5\arcsec$ reaches $8\cdot10^{-3}$ and almost does not change throughout the object. A value of $f_{\rm C}^{\rm bPAH}$ decreases with time, and it is lower than $f_{\rm C}^{{\rm C}_{60}}$ and $f_{\rm C}^{{\rm C}_{70}}$. It reflects the fact that the PAHs, which turn to bPAHs, have already completed this transformation by 10$^4$~yr, and these bPAHs have already mostly turned to fullerenes. On the left panel, $f_{\rm C}^{\rm PAHs}$ is much lower than on the right panel, i.e. account of hydrogenation significantly reduces the abundance of PAHs due to their rapid hydrogenation and subsequent destruction of HPAHs. Even though this represents an extreme case, it is worth considering. At 10$^4$~yr $f_{\rm C}^{\rm PAHs}$ is lower than $f_{\rm C}^{{\rm C}_{60}}$ and $f_{\rm C}^{{\rm C}_{70}}$ at the distance $\geqslant7-8\arcsec$, but at 10$^5$~yr PAHs are destroyed even farther from the star and attain abundances comparable to that of C$_{60}$ only at the distance $\gtrsim17\arcsec$. Thus, in this case, PAHs and fullerenes are separated spatially, fullerenes dominate at the closest vicinity of the star, and PAHs dominate farther away.

The value of $f_{\rm C}^{{\rm C}_{70}}$ is even higher than $f_{\rm C}^{{\rm C}_{60}}$ at a distance of $\lesssim 10\arcsec$. It reflects the fact that the abundance of C$_{70}$ is compensated by dissociation of larger PAHs with $N_{\rm C}>78$, which gradually evolve to smaller PAHs, dPAHs and bPAHs until they are stalled in the C$_{70}$ bin. Thus, the bin of C$_{70}$ accumulate the products of evolution of PAHs with larger $N_{\rm C}$. Then C$_{70}$ can shrink to smaller fullerenes, and with time the bin of C$_{60}$ will accumulate products of evolution of C$_{70}$. But it takes longer than 10$^5$~yr.

\begin{figure}
\includegraphics[width=0.45\textwidth]{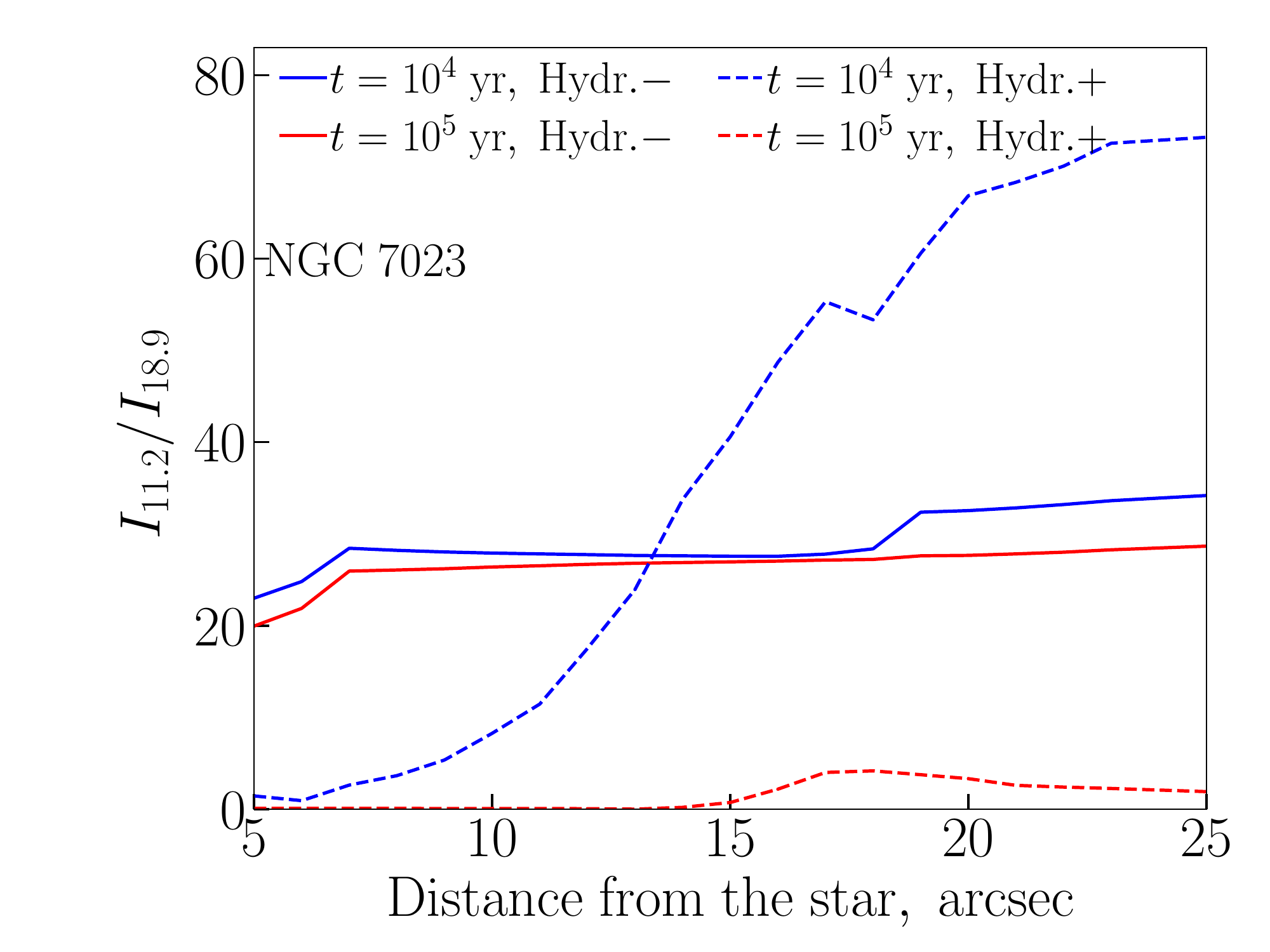}
\caption{The dependence of $I_{11.2}/I_{18.9}$ on the distance from the star. The solid lines correspond to the modelling without hydrogenation, the dashed lines --- with included hydrogenation.} 
\label{fig: ratio_ngc7023}
\end{figure}

We present the ratio of intensities at 11.2 and 18.9~$\mu$m bands along NGC~7023 in Fig.~\ref{fig: ratio_ngc7023} analogously to the Orion Bar. We demonstrate the results when the hydrogenation is included (solid lines) and neglected (dashed lines). The ratio grows as the distance increases. It varies from $\approx20$ to $\approx35$ at 10$^5$~yr if the hydrogenation is not included in the modelling and from $\approx0.1$ to $\approx4$ if the hydrogenation is accounted for. At 10$^4$~yr the ratio is higher in both cases. The prominent growth of the ratio is seen at this age if the hydrogenation is included (dashed blue line). It reflects the behaviour of $f_{\rm C}^{\rm PAHs}$ in Fig.~\ref{fig: fc_ngc7023} (left panel), where at 10$^4$~yr PAHs are still not destroyed as much as at 10$^5$~yr. PAHs dominate over fullerenes at a distance of $\geqslant7-8\arcsec$. The farther from the star, the more PAHs are abundant. Consequently, the ratio grows and may achieve values up to $\approx70$. At 10$^5$~yr PAHs are mostly destroyed, and, consequently, their emission is much weaker than the fullerene emission. Note that the value of the ratio is lower than in the Orion Bar. Only at $A_{\rm V}=0.1$ in the Orion Bar the ratio is comparable to the ratio in NGC~7023.

\subsection{Grid of parameters}
\label{sec: grid}

In this section, we present the results of modelling of fullerene evolution on the grid of parameters, which is supposed to cover a major part of the ISM conditions. The hydrogenation is also included to trace the maximum amount of fulleranes. Overall there are four key parameters affecting the PAH evolution, namely, $U$, $T_{\rm gas}$, $n({\rm H})$, and $n_{\rm e}$. 

\begin{figure}
\includegraphics[width=0.45\textwidth]{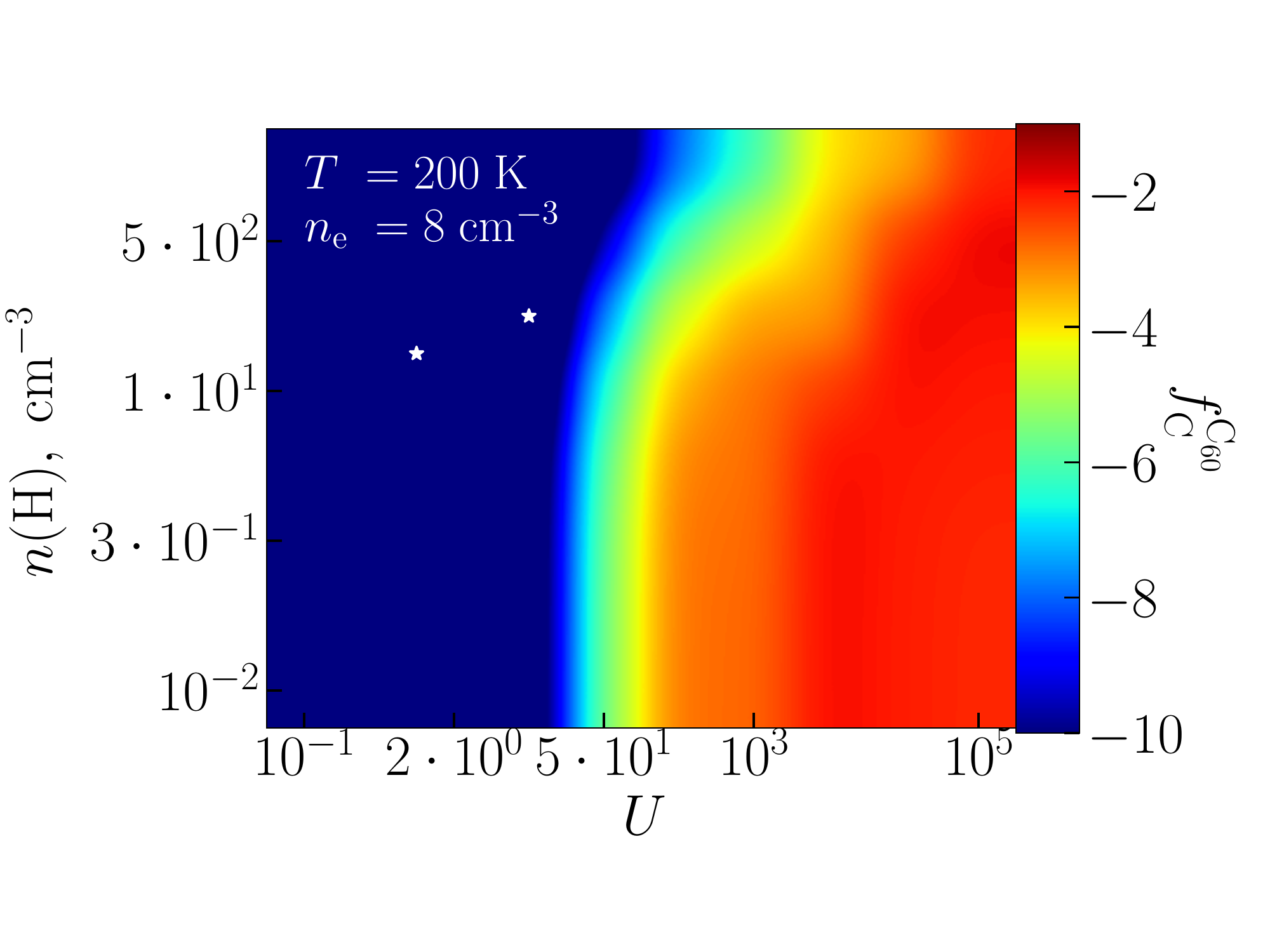}
\caption{The fraction of C locked in C$_{60}$ as a function of $n({\rm H})$ and $U$. Temperature and electron density are fixed. The hydrogenation is included. White stars show average conditions taken from \citet{falgarone05,ingalls11} for the lines of sight along which C$_{60}$ was detected.}
\label{fig: fc_map}
\end{figure}

In Fig.~\ref{fig: fc_map} we demonstrate a dependence of carbon fraction locked in C$_{60}$ on $U$ and $n({\rm H})$ with fixed $T_{\rm gas}=200$~K and $n_{\rm e}=8$~cm$^{-3}$. Colour indicates the fraction of C locked in C$_{60}$. It is seen that the favourable conditions for the fullerene formation correspond to high radiation field intensity, and their formation becomes less efficient as the hydrogen density increases. The highest fraction obtained in this study is around 0.01, i.e. around 1\% of C can be locked in C$_{60}$. Note that the efficient fullerene formation corresponds to $U \gtrsim 10^{3}$, but if $n({\rm H})\gtrsim 10^{1}-10^{2}$~cm$^{-3}$, then $f_{\rm C}^{{\rm C}_{60}}$ is lower at the same $U$.

\begin{figure}
\includegraphics[width=0.45\textwidth]{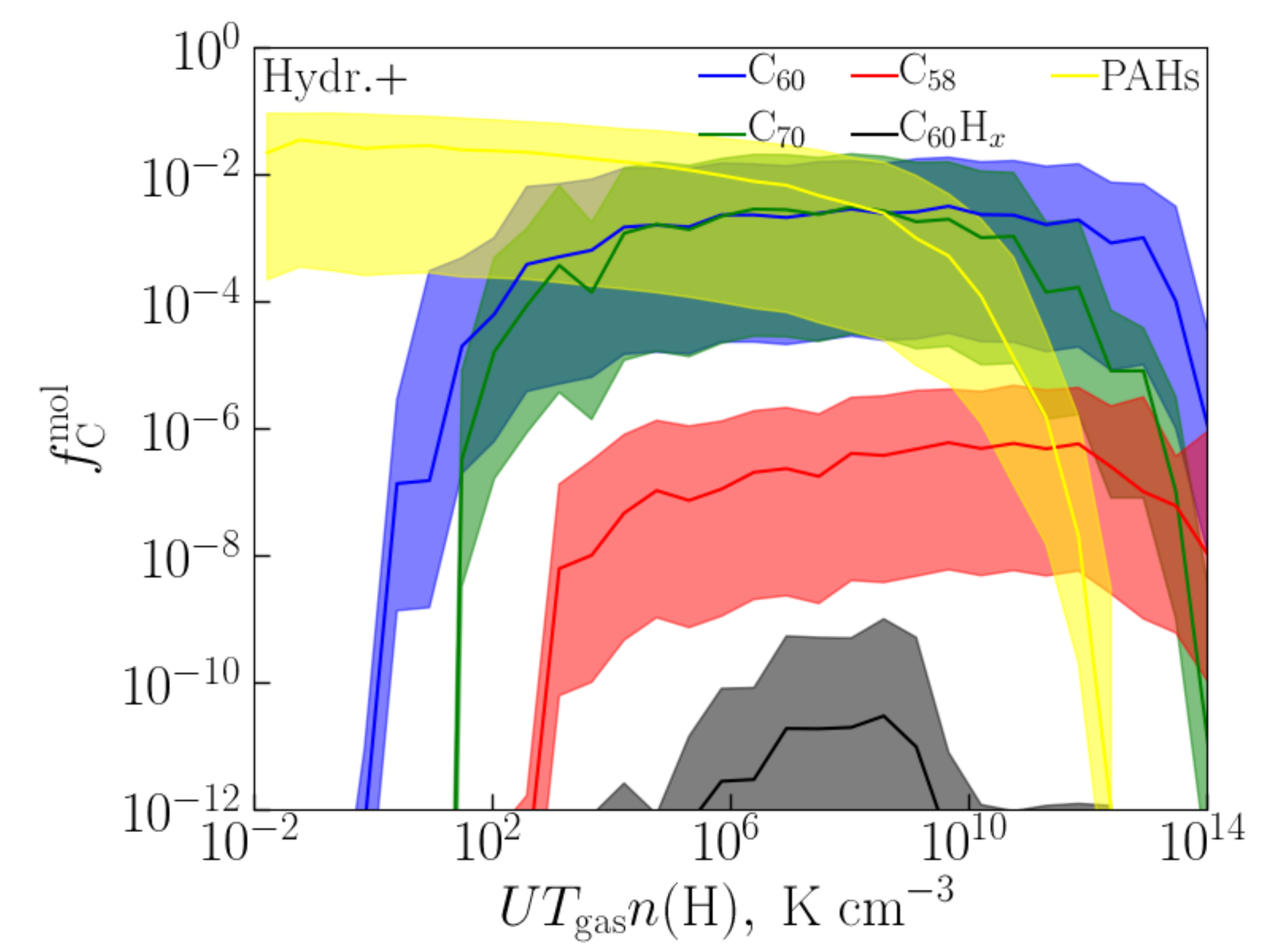}
\caption{The fraction of C locked in C$_{60}$, C$_{70}$, C$_{58}$, and C$_{60}$H$_{x}$ fullerenes as well as planar PAHs versus the parameter $U T_{\rm gas}n({\rm H})$. The hydrogenation is included.}
\label{fig: fc_grid}
\end{figure}

In Fig.~\ref{fig: fc_grid} we show the dependence of $f_{\rm C}^{{\rm C}_{60}}$, $f_{\rm C}^{{\rm C}_{70}}$, $f_{\rm C}^{{\rm C}_{58}}$, $f_{\rm C}^{{\rm C}_{60}{\rm H}_{x}}$ on $U T_{\rm gas} n({\rm H})$. We found that if hydrogenation is included, this parameter correlates with fullerene abundance better than $U/n({\rm H})$. The fractions, presented in Fig.~\ref{fig: fc_grid}, can be considered as upper limits for PDRs because: 1) the results are obtained with the Laplace method, which was found to provide the most optimistic description of the fullerene formation (see Murga et al. 2022 in prep.); 2) the results correspond to $10^6$~yr.

We vary the four parameters, and the results are sensitive to all of them to a certain degree. However, we found that the general behaviour of this sensitivity is well characterised by the value of $U T_{\rm gas} n({\rm H})$. Varying all the parameters, for each specific value of $U T_{\rm gas} n({\rm H})$ we obtain a vertical distribution of points on a diagram, relating $U T_{\rm gas} n({\rm H})$ to the fraction of carbon atoms locked in a certain species. In Fig.~\ref{fig: fc_grid} we illustrate such distributions for $f_{\rm C}^{{\rm C}_{60}}$, $f_{\rm C}^{{\rm C}_{70}}$, $f_{\rm C}^{{\rm C}_{58}}$, and $f_{\rm C}^{{\rm C}_{60}{\rm H}_{x}}$. Solid lines show loci of the average fractions, while the colour corridors represent three standard deviations for the corresponding fraction. When $U T_{\rm gas}n({\rm H})$ is lower than $10^2-10^3$~K~cm$^{-3}$, i.e. the medium is relatively cold and the radiation intensity is relatively low, fullerenes are formed slowly, while PAHs are abundant. When the parameter $U T_{\rm gas} n({\rm H})$ is higher than $10^2-10^3$~K~cm$^{-3}$, the efficiency of the fullerene formation increases and fractions of fullerenes gradually reach the plateau, when $U T_{\rm gas} n({\rm H})\approx10^4$~K~cm$^{-3}$. The value of the plateau is limited by the amount of fullerene-producing PAHs. In the opposite, the PAH fraction gradually decreases until it drops quickly, when $U T_{\rm gas} n({\rm H})$ exceeds $\approx10^{2}$~K~cm$^{-3}$. The fractions of C$_{60}$, C$_{70}$, C$_{58}$ remain high up to $U T_{\rm gas} n({\rm H})\approx10^{10}-10^{12}$~K~cm$^{-3}$. At the very high values of $U T_{\rm gas} n({\rm H})$ the fractions of all fullerenes decrease, but the fraction of C$_{58}$ decreases less steeply than C$_{60}$ and C$_{70}$, as C$_{60}$ and C$_{70}$ are gradually dissociated and turn to C$_{58}$. Some fraction of fulleranes may appear at $10^4<U T_{\rm gas} n({\rm H})<10^{10}$~K~cm$^{-3}$ although their abundance is low. We note that here the corridors do not cover all possible solutions, e.g. if $U<10^2-10^3$, $f_{\rm C}^{{\rm C}_{60}}$ is much lower than the bottom of the illustrated corridor. Thus, the shown relation works only for the enhanced radiation field.

The results presented in Fig.~\ref{fig: fc_grid} point out a general trend that the fullerene formation is the most efficient, when the product of the radiation field intensity, temperature and hydrogen density is within a wide range, from 10$^{2}$ to $10^{13}$~K~cm$^{-3}$. The environment can be either H-poor with the intense radiation field or H-rich with the modest radiation field, herewith $U$ must be higher than $10^2-10^3$. However, if all three parameters are large, the formation of fullerenes is suppressed as well as if all three parameters are low. There is also a significant scatter in the final fullerene abundance depending on all the parameters.

\section{Discussion}

\subsection{Comparison with observations}

\subsubsection{The Orion Bar PDR}

\cite{salgado16} presented SOFIA photometric profiles of the Orion Bar at different wavelengths including 7.7 and 19~$\mu$m, which can be attributed to PAHs and fullerenes, respectively. According to their results, the PAH and fullerene emission peaks almost coincide and are located at around $117.5-118\arcsec$. The ionisation front traced by the {O\,{\sc{i}}~}6300~\AA{} line is located closer to the ionising star at around $113\arcsec$. Our results obtained with the model with hydrogenation indicate that the emission peaks of PAHs and fullerenes should be shifted from the IF towards the molecular cloud. The peaks are around $A_{\rm V}\approx0.9-1.0$. Although, if the hydrogenation is not included in the model, the peaks are exactly at the IF. As we do not model the dynamics of the PDR, we cannot restore the precise position of emission peaks. Nevertheless, the modelled shift of the PAH and fullerene emission peaks is also seen in the observations, thus, this shift can be caused by the rapid destruction of PAHs, which are saturated by hydrogen atoms there. However, we admit that the observed distance between the IF and PAH and fullerene emission peaks is around $4-5\arcsec$, and it is probably closer than the modelled peak position at $A_{\rm V}\approx0.9-1.0$. A combination of two our models by fine tuning the hydrogenation efficiency likely could describe the observations more precisely.  

\cite{boersma12} presented mid-IR spectroscopic observations of the Orion Bar, which cover the distances from $\approx2.5$ to $12\arcmin$. We note again that following \citet{goicoechea15}, we express a distance in $A_{\rm V}$ units in our model of this object, therefore, we cannot compare our modelling results with the observational results directly, although some juxtaposition of trends can be done. \cite{boersma12} showed that the PAH-to-fullerene ratio increases with distance, however, a trough in this ratio is seen between $8\arcmin$ and $10\arcmin$, where the medium is shielded and molecular. Then, the ratio increases again. \cite{boersma12} divided their observations into three parts: 1)~{H\,{\sc{ii}}~}--PDR interface, 2)~shielded region, and 3)~outermost {H\,{\sc{ii}}~}--PDR interface with limb brightening of PAH emission. Our modelling in the range of $A_{\rm V}$ from 0.1 to $\approx1.5$ supposedly corresponds to the first part of the observations (namely, locations I1--M4 defined in \cite{boersma12}, although the observations in locations I1--I4 have quite large uncertainties). The growth of the PAH-to-fullerene ratio with the distance can be well described by our modelling, i.e. this trend reflects fullerene formation from PAHs by the top-down scenario under harsh radiation field in this object. The parameter $U T_{\rm gas} n({\rm H})$ defined in Sect.~\ref{sec: grid} changes dramatically across the Orion Bar, from 10$^{13}$~cm$^{-3}$~K at $A_{\rm V}=0.1$ to 10$^{-2}$~cm$^{-3}$~K at $A_{\rm V}=10$ and almost covers the considered range of the parameter in Fig.~\ref{fig: fc_grid}. So, in the Orion Bar we encounter all the conditions, depicted in Fig.~\ref{fig: fc_grid}, from the region, where fullerenes dominate over PAHs at $A_{\rm V}=0.1$ to the region, where PAHs dominate, and fullerenes are not formed from PAHs.

\cite{castellanos14} estimated $f_{\rm C}^{\rm C_{60}}$ and $f_{\rm C}^{\rm PAHs}$ in the Orion Bar to be $\approx 5\cdot 10^{-4}$ and 0.01, respectively. Our modelled estimations of $f_{\rm C}^{\rm PAHs}$ are consistent with these values, $f_{\rm C}^{\rm C_{60}}$ is consistent at $1.5<A_{\rm V}<2.5$, while at the locations closer to the stars the modelled $f_{\rm C}^{\rm C_{60}}$ is higher and reaches 7$\cdot10^{-3}$ (if the hydrogenation is considered, then $f_{\rm C}^{\rm C_{60}}$ drops near the IF). According to the adopted initial dust size distribution taken from \citet{wd01} the fractions of C$_{66}$H$_{20}$ and C$_{78}$H$_{22}$ are slightly less than 1\%. Thus, at $A_{\rm V}\approx 0.1$ all C$_{66}$H$_{20}$ and C$_{78}$H$_{22}$ convert to fullerenes C$_{60}$ and C$_{70}$, correspondingly. Value of $f_{\rm C}^{\rm C_{70}}$ drops very quickly as $A_{\rm V}$ increases, while $f_{\rm C}^{\rm C_{60}}$ stays nearly constant up to $A_{\rm V}\approx1.5$, and then also drops very quickly. The plateau of $f_{\rm C}^{\rm C_{60}}$ at $0.1<A_{\rm V}\approx1.5$ (if the hydrogenation is not included) reflects the maximum of fullerenes that can be formed from PAHs with the adopted distribution and under the considered conditions. Thus, the modelled estimations of $f_{\rm C}^{\rm C_{60}}$ can explain the estimates of \citet{castellanos14} based on the observations of \citet{boersma12}.

The ratio between surface brightnesses of the bands at 11.2 and 18.9~$\mu$m, $I_{11.2}/I_{18.9}$, increases from $\approx5$ to $\approx27$ according to \cite{boersma12}. The ratio calculated based on our calculations (Fig.~\ref{fig: ratio_av_orion_laplace}) varies from $\approx25$ to $\approx2000$. Thus, the modelled ratio corresponding to the brightest location of the Orion Bar ($A_{\rm V}=0.1$) is comparable with the observational ratio. We suppose that modelled region at $A_{\rm V}=0.1-1.5$ approximately corresponds to I1--M4 locations in \cite{boersma12}, and, therefore, our ratio is less by at 2--8 times at these locations. At other locations our ratio becomes less by 1--2 orders of magnitude. Note that we discuss the results without hydrogenation, i.e. where the fullerene abundance does not drop at $A_{\rm V}=0.1$. If we include the hydrogenation, our modelled estimates become even lower. 

The underestimation of the ratio can be explained by several reasons. First, the observed fullerenes might have been partially formed somewhere else in the ISM via some alternative mechanism and were brought into the Orion Bar PDR. Some fraction of fullerenes could be formed from PAHs due to the mechanism considered in this work, and the contribution of these fullerenes changes the profile of the PAH-to-fullerene ratio, but only within a narrow distance range near the IF. At other locations, for example, at the outer edge of the molecular cloud, this contribution is predicted to be negligible and cannot influence the PAH-to-fullerene ratio, while fullerenes are observed there. \cite{boersma12} adhere this point of view as well, and they considered the anomalous trough of the PAH-to-fullerene ratio in the shielded region as a support of this point. Second, a clumping structure of the Orion Bar~\citep{goicoechea16} can be a reason, i.e. UV photons penetrate deeper, and, consequently, UV radiation field intensity is actually higher in the molecular cloud than it is adopted in the model. Thus, fullerenes can form deeper in the cloud.

The third reason of the theoretical underestimation of the ratio is possible overestimation of the intensity of the 11.2~$\mu$m band. PAHs in the Orion Bar are rather dehydrogenated. The band at 11.2~$\mu$m is believed to arise from bending of solo CH bonds~\citep{hudgins99}. This band is affected by the PAH hydrogenation state according to the theoretical calculations provided by the NASA Ames PAH IR Database. The absorption cross-section suggested in the model of \cite{DL07} does not depend on the hydrogenation state. We implemented some modifications related to the hydrogenated state in our previous work \citep{murga_orion} and applied them in this study. However, these modifications are quite uncertain, and it is possible that the actual 11.2~$\mu$m band is even weaker that our modifications imply.

The fourth reason can be hidden in the adopted initial dust size distribution taken from the WD01 model. It assumes the certain amount of PAHs of each size relative to hydrogen. Although the WD01 model generally fits observations of the ISM, it does not necessary suit the Orion Bar with the physical conditions, which are far from the conditions of the diffuse ISM. Thus, the abundance of PAHs, which are able to turn to fullerenes can indeed be larger than our modelling suggests, and, therefore, the larger amount of fullerenes can be formed due to PAH dissociation. 

\subsubsection{NGC~7023}

Our modelling demonstrates that without hydrogenation the PAH abundance modestly varies along the PDR NGC~7023 and is lower than the initial value ($\approx 7\cdot10^{-2}$) by at most around 20\%. The fullerene (at least C$_{60}$) abundance remains almost unchanged. If the hydrogenation is included in the modelling, than the PAH abundance varies substantially: after $10^5$~yr PAHs are mostly destroyed at $5-10\arcsec$. Their abundance increases only at $20-25\arcsec$, but is still lower than the initial abundance by about 2 orders of magnitude. According to the observations presented in \cite{sellgren10} the fullerene emission has its maximum near the star and decreases at $20-25\arcsec$, where the PAH emission reaches its maximum, i.e. the peaks of the PAH and fullerene emission are spatially separated unlike in the Orion Bar. This behaviour is close to our modelled results with the Laplace method. The maxima are more prominent if the hydrogenation is included.

According to \cite{berne15}, $f_{\rm C}^{{\rm C}_{60}}$ should be in the range $\approx 3\cdot10^{-5}-3\cdot 10^{-3}$. In our modelling case with the maximum fullerene yield (with included hydrogenation), values of $f_{\rm C}^{{\rm C}_{60}}$ and $f_{\rm C}^{{\rm C}_{70}}$ may reach $10^{-2}$. Thus, our estimates of $f_{\rm C}^{{\rm C}_{60}}$ are at least three times higher than the estimates of \cite{berne15}. We emphasise that in this case PAHs with $N_{\rm C} \geqslant 66$ subsequently turn to fullerene-producing PAHs and further to fullerenes under the adopted conditions of NGC~7023, so the final $f_{\rm C}^{{\rm C}_{60}}$ and $f_{\rm C}^{{\rm C}_{70}}$ are even slightly higher than the initial values of $f_{\rm C}^{{\rm C}_{66}{\rm H}_{20}}$ and $f_{\rm C}^{{\rm C}_{78}{\rm H}_{22}}$. Our result should be considered as an upper limit of fullerene abundance. Nevertheless, in this object the needed amount of fullerenes can be explained by the formation from PAHs only.  

\cite{berne15} obtained that C$_{66}$H$_{20}$ totally disappears by $10^2$~yr at $5-25\arcsec$. At 10$^5$~yr C$_{60}$ has its maximum abundance at $10\arcsec$, while at $5\arcsec$ C$_{60}$ is already dissociated to C$_{58}$ by this time. The abundance of C$_{58}$ in our modelling is lower by $\sim$3 orders of magnitude than the abundance of C$_{60}$ even at $5\arcsec$. The most significant difference in our calculation methods is a set of vibrational frequencies for C$_{60}$. \cite{berne15} used the frequencies from \cite{schettino01}, while we used more recent data provided by the NASA Ames PAH IR database. This set of frequencies is broader than the set from \cite{schettino01}. The dissociation rate calculated with the Laplace method is sensitive to the number of frequencies, and it decreases as this number grows. Therefore, the dissociation rate of C$_{60}$ in our work is lower than the rate in \cite{berne15}. Thus, C$_{60}$ does not dissociate to C$_{58}$ by 10$^{5}$~yr. The abundance of C$_{60}$, computed by \cite{berne15}, gradually decreases with the distance from the star, and this result is consistent with our modelling.

We took the spectra presented in \cite{sellgren10} and \cite{berne12} and measured the fluxes in the bands at 11.2 and 18.9 $\mu$m in the same way as it was performed for synthetic spectra. We found that the ratio $F_{11.2}/F_{18.9}$ is around 15 for the spectrum presented in \cite{berne12} and around four for the spectrum presented in \cite{sellgren10}. The ratio based on our synthetic spectra varies from $\approx0.1$ to 4 if the hydrogenation is included and from $\approx20$ to 35 if there is no hydrogenation. Also the ratio differs at 10$^{4}$ and 10$^5$~yr. The observational ratios fall into the range between our two border estimates when the destruction is extremely rapid due to presence of unstable PAHs in hydrogenated states and when the destruction occurs modestly. The age of NGC~7023 is estimated to be from 10$^{4}$ to 10$^5$~yr, and our synthetic ratio for these two ages also represent borders within which the observational values fall. The parameters of most considered processes are just assumptions as neither measurements nor calculations are available for them to date. So, we suppose that experimental or theoretical refinement of these parameters will lead to the better consistency between our estimates and observations.

\subsection{Grid of parameters}

We presented the results of modelling of the fullerene evolution in a wide range of parameters. We defined the parameter $U T_{\rm gas} n({\rm H})$, which characterises the fullerene formation efficiency. We showed that the fullerene formation due to the PAH dissociation is efficient when the parameter $U T_{\rm gas}n({\rm H})$ is between $10^2$ and $10^{10}$~cm$^{-3}$~K. However, the scatter in the final fraction of fullerenes is quite large. At the same value of $U \sqrt{T_{\rm gas}} n({\rm H})$, the modelled values of $f_{\rm C}^{{\rm C}_{60}}$, $f_{\rm C}^{{\rm C}_{70}}$, and also $f_{\rm C}^{\rm PAHs}$ can differ by several orders. The knowledge of all parameters ($T_{\rm gas}$, $n({\rm H})$, $n_{\rm e}$, $U$) is required for more accurate prediction of $f_{\rm C}^{{\rm C}_{60}}$ and other fractions. Additionally, the age is also an important factor as all the fractions change with time. 

\cite{campbell15} put together the observational data of different PDRs including the Orion Bar and NGC~7023. They showed that there is no a common trend between $f_{\rm C}^{{\rm C}_{60}}$ and $G_0$ or $G_0/n({\rm H})$ for all the objects taken together, although trends can be seen for individual objects. Based on our modelling, we conclude that, firstly, the lack of the common trend may be related with differences in individual physical parameters and age, and, secondly, some fraction of fullerenes could be formed elsewhere in the ISM and brought into observed PDRs. Thus, fullerenes can exist in all PDRs, and their abundances may increase due to the formation via PAH dissociation differently depending on the conditions. We conclude that the fullerenes, which are observed at the closest locations to the ionising source of the Orion Bar and near the star in NGC~7023, could have been formed mostly from PAH dissociation, while fullerenes at the locations far from the ionising source of the Orion Bar have been brought from the ISM. Thus, PAH dissociation can be an efficient mechanism of fullerene formation only in the closest vicinity of a strong UV radiation source.

It is believed that the top-down mechanism (formation of fullerene from PAHs) may work only in H-poor environment, however, we found that modest hydrogen number density up to $10^3-10^4$~cm$^{-3}$ does not lead to substantial suppression of fullerene formation. Only at $n({\rm H})>10^4$~cm$^{-3}$ the fullerene formation indeed subsides. As for NGC~7023 with $n({\rm H})\approx10^3$~cm$^{-3}$, fullerenes can be efficiently formed from PAH dissociation. However, we suppose that there is another factor that influences the efficiency of the top-down scenario. If the parameters considered in this work (the radiation field intensity, hydrogen number density, and gas temperature) were the only parameters that influence on fullerene formation, fullerenes would be found in a larger number of planetary nebulae than they were indeed found~\citep{otsuka14}. Apparently, some other factors (e.g. metallicity) play a noticeable role in this process.

Fullerenes were detected in the diffuse ISM both in emission and absorption~\citep{berne17}. We took the parameters along the lines of sight from \cite{falgarone05, ingalls11} and averaged them. These averaged parameters are shown in Fig.~\ref{fig: fc_map} with white stars. It is seen that fullerenes cannot be formed under conditions of the diffuse ISM as their formation requires more energetic environments. The fullerenes in diffuse clouds were likely brought there from some other parts of the ISM.

\subsection{Notes on some molecules}
\subsubsection{C$_{70}$}

C$_{70}$ was identified along with C$_{60}$ in the planetary nebula Tc~1~\citep{cami10}. Its abundance was found to be comparable with the abundance of C$_{60}$. On the other hand, \cite{garcia11} found that the ratio of C$_{70}$ and C$_{60}$ abundances is 0.02--0.2 in planetary nebulae of Magellanic Clouds. \cite{omont16} mentioned that the abundance of C$_{70}$ can be underestimated due to its efficient cooling by Poincare fluorescence which is not so important for C$_{60}$. Thus, C$_{70}$ likely presents in many objects along with C$_{60}$ and its actual abundance can be higher than it has been estimated so far. Our results show that the abundance of C$_{70}$ is mostly lower than the abundance of C$_{60}$, although C$_{70}$ can dominate under conditions with extremely high UV radiation like at the closest locations to the star in NGC~7023.

\cite{garcia11} supposed that the ratio between abundances of C$_{70}$ and C$_{60}$ may reveal the formation scenario because the ratio varies depending on the production method~\citep{mansurov11}. This ratio is around 0.02--0.18 in vapour condensation scenario~\citep{mansurov11}. The value of the ratio for the top-down scenario is undetermined in laboratory. Based on our results we can conclude that the ratio varies from 0 to several times depending on physical conditions and object's age. However, we obtained that in NGC~7023 C$_{70}$ may dominate over C$_{60}$ in the locations, closest to the star, where fullerene production (both C$_{60}$ and C$_{70}$) is the most efficient. In Fig.~\ref{fig: fc_grid} it is also seen that $f_{\rm C}^{{\rm C}_{60}}$ and $f_{\rm C}^{{\rm C}_{70}}$ are comparable, when conditions are favourable for fullerene formation. Thus, it can be concluded that the ratio between abundances of C$_{70}$ and C$_{60}$ of around unity points on their formation by the top-down scenario in the object, where they are observed, in addition to other possible and more traditional scenarios of formation in stellar outflows, where this ratio cannot achieve such high values. 

\subsubsection{Fulleranes}

In spite of intense search of fulleranes in the ISM and their potential significance for explanations of some observational features~\citep{iglesias05, iglesias12, cataldo14}, our modelling shows that they may present in the ISM only in very low abundance. The fraction of carbon locked in fulleranes does not exceed 10$^{-9}$, which is lower by 7--8 orders of magnitude than in pure fullerenes. Their observational detection is thus complicated.

\subsubsection{C$_{60}$, C$_{60}^{+}$ and C$_{70}$ in IC~348}

Our results can also be compared to observations of IC~348, where C$_{60}$, C$_{60}^{+}$ and C$_{70}$ have been detected by \cite{iglesias19}. Thus, we can use the abundance estimates presented in this work as a test for the abundance ratios of these fullerenes obtained with our model. IC~348 is a large star-forming region with a number of young stars. \cite{iglesias19} obtained estimates of fullerene abundance around three stars, LRLL~1 (B type), LRLL~2 (A type), and LRLL~58 (M type). The radiation field around these stars is not as intense as in the Orion Bar and NGC~7023. \cite{iglesias19} adopted that the radiation field intensity expressed in $G_0$ (which is $\approx1.14U$) is about 45 for LRLL~1, and about 20 for the other two stars. The fraction of carbon locked in C$_{60}$ was estimated to be 1.6$\times10^{-3}$, 0.8$\times10^{-3}$ and 0.8$\times10^{-3}$ for LRLL~1, LRLL~2 and LRLL~58, respectively. According to our calculations, e.g. presented in Fig.~\ref{fig: fc_map}, the formation of fullerenes via the top-down mechanism is not efficient if the radiation field intensity is lower than $10^2-10^3$, therefore, the values of $f_{\rm C}^{{\rm C}_{60}}$ cannot be achieved. However, \cite{iglesias19} mentioned that the adopted radiation field is an average one, and the actual intensity can vary by few orders within the considered region. If we assume that the local intensity is around $5\times10^2-10^3$ or higher, and the number density of atomic hydrogen does not exceed $10^2-10^3$~cm$^{-3}$ than the modelled $f_{\rm C}^{{\rm C}_{60}}$ are consistent with observational estimates.

In the Orion Bar such radiation field ($U\approx 500$), somewhat enhanced relative to the average field in IC~348, corresponds to $A_{\rm V}\approx 2$, so we suppose that the results obtained for this location can be relevant for the observations of IC~348. At $A_{\rm V}\approx 2$ neutral C$_{60}$ dominate having the fractional abundance around 55\%, the fraction of C$_{60}^{+}$ is around 10\%, while anions C$_{60}^{-}$ have the fraction around 35\%. Thus, the ratio of the C$_{60}^{+}$ abundance to the C$_{60}$ abundance is $\approx 0.15-0.2$. The observational value of the ratio is $\approx0.1$. Taking into account that the actual parameters of the medium, which contains the fullerenes, are unknown, and adopted parameters are quite rough, we suppose that the modelling results satisfactory describe the observed ionisation state. 

Further, \cite{iglesias19} showed that the C$_{70}$ abundance is around a half of the C$_{60}$ abundance near the star LRLL~2 and equals to the C$_{60}$ abundance near the star LRLL~58. These abundances are quite large. In Fig.~\ref{fig: fc_grid}, $f_{\rm C}^{{\rm C}_{70}}$ can be comparable with $f_{\rm C}^{{\rm C}_{60}}$ withing the range of $U \sqrt{T_{\rm gas}} n({\rm H})$ from $10^4$ to $10^{10}$~cm$^{-3}$~K. However, these high values can be achieved if the radiation field intensity is high ($>10^3$) and the atomic hydrogen density is low ($<10^{3}$~cm$^{-3}$). The stars LRLL~2 and LRLL~58 are relatively cold, and high UV radiation is not expected there. It cannot be excluded that there can be some locations in IC~348 where the required conditions were satisfied and C$_{70}$ was formed, but it could have hardly been formed in situ near these stars. It is more likely that C$_{70}$ fullerenes were formed elsewhere and were brought here.

\subsection{Alternative routes for fullerene formation}

Fullerenes were firstly produced in hot carbon vapour in experiments \citep{kroto85, kratschmer90, smalley92}. Such methods suppose the bottom-up scenario when fullerenes grow via subsequent clusterisation of C$_2$, although the exact way of their clusterisation is debated~\citep{smalley92, Lozovik97, dunk12}. Recently \cite{dunk12} demonstrated the model of closed network growth to describe the clusterisation. This bottom-up scenario suits to conditions of stellar outflows or supernovae~\citep{cherchneff00, jager09, dunk12} and was considered to be efficient only for H-poor environment, while fullerenes are detected, rather in H-rich objects~\citep{garcia10, garcia11}. However, \cite{dunk12, dunk13} experimentally showed that the process occurs in H-rich environment as well and, therefore, can be considered as a valuable source of fullerenes. 

Besides stellar outflows, fullerenes can be formed by the bottom-up route during the dust shattering due to grain-grain collisions~\citep{omont16}. The collisions may occur with high velocities (up to hundreds of km/s) in shocks. Such hard collisions inevitably lead to melting and evaporation of dust grains. The carbon vapour that appears due to the collisions has high temperature and pressure, and, what is important, rapidly cools down after the grain destruction. These conditions are very similar to those that are tuned to form fullerenes. The cooling carbon vapour condensates and may form different carbon clusters including fullerenes. This formation route can be efficient in highly turbulent medium that arises after multiple shocks. One needs to perform precise calculations to estimate the contribution of this route of fullerene formation.

Besides the `classical' formation, other routes have been proposed and studied in laboratory. One of them is a top-down PAHs photo-processing considered in this work and previously in \cite{berne12, zhen14_ful, berne15}. Another variant of the top-down scenario is photo-processing of hydrogenated amorphous carbon (HAC) grains~\citep{scott97, garcia11, mic12}. Fullerenes are indeed experimentally detected as products of HAC dissociation~\citep{scott97, gadallah11, duley12}. However, experiments mostly use hydrogenated `soft' HACs, while modern interstellar dust models (e.g. the one presented in \citealt{jones13}) suggest that hydrogenated amorphous carbons, being one of the major constituents of interstellar carbonaceous dust, are characterised by the low fraction of hydrogen~\citep{chiar13, ld12}. According to the model of \cite{jones13}, the fraction of hydrogen in small dust grains or on the surface of large grains must be just around 0.1. Such a material can become resistant to UV irradiation, so the results of the experiments with `soft' HAC grains should be supplemented with additional estimates of the photo-processing efficiency for `hard' HAC grains. Additionally \cite{duley12} suggested that we observe `proper' fullerenes only in objects with extreme radiation such as PN Tc~1, while in other objects we observe proto-fullerenes, i.e. a stage between a HAC grain and a fullerene.

Besides photons, highly energetic particles in shock waves may dissociate PAHs or HACs that may lead to fullerene formation. This route of fullerene formation was considered in experiments of \cite{chuvilin10} and dynamical simulations of \cite{sinitsa17}, although application of these works to astrophysical objects has not been done yet. However, 
\cite{bernal19} suggested that bombardment of crystalline silicon carbide grains could be a possible way for explanation of fullerene abundance in the ISM, but the question about abundance of these grains themselves in the ISM is debatable. 

\subsection{Uncertainties and topics for further investigations}

Our model does not include some processes relevant for PAH and fullerene evolution. Among them are growth of PAHs through reactions with gaseous carbons and small hydrocarbons, formation of PAH and fullerene dyads, endofullerenes, dopped fullerenes and fullerenes of wide range of size and also nanotubes, which are in fact similar to fullerenes, etc. It is likely that some forms of PAHs and fullerenes may explain some unidentified features including diffuse interstellar bands. 

Our results indicate the importance of hydrogenation because inclusion of this process into the model allows explaining some observational features that cannot be reproduced without hydrogenation. On the other hand, in some cases, hydrogenation leads to extremely rapid dissociation of PAHs and underestimation of their abundance. Thus, the rates of hydrogenation of PAHs should be re-estimated. Dissociation rates also need to be specified. This especially concerns large PAHs and HPAHs.

Laboratory measurements of the physical and chemical properties of fullerenes, fulleranes, dPAHs, bPAHs such as absorption cross-section over wider wavelength range, critical energy required for their dissociation, vibrational properties are needed.  

\section{Summary}\label{sect: summary}

We developed the model of evolution of PAHs and fullerenes in a UV irradiated environment. The model includes photo-dissociation of PAHs, hydrogenation of PAHs, their isomerisation to fullerenes, photo-dissociation of fullerenes, hydrogenation of fullerenes. We considered the modelling with hydrogenation and without it and discussed how this process influences the results. We modelled abundance of PAHs of different sizes and fullerenes from C$_{58}$ to C$_{70}$, taking into account their charge and hydrogenation state. We performed calculations for two PDRs, the Orion Bar and NGC~7023, and also for a grid of parameters: radiation field intensity, gas temperature, hydrogen and electron density. We estimated the fraction of carbon locked in PAHs, C$_{60}$, and C$_{70}$ depending on conditions. We calculated the IR emission spectra of modelled PAH and fullerene compositions in the Orion Bar and NGC~7023 and ratio between the PAH band at 11.2~$\mu$m and the fullerene band at 18.9~$\mu$m along the PDRs. 

We found that if the hydrogenation is not included, in the Orion Bar $f_{\rm C}^{{\rm C}_{60}}$ may be up to 7$\cdot10^{-3}$ at $A_{\rm V}=0.1-1$ and $f_{\rm C}^{{\rm C}_{70}}$ may be up to 8$\cdot10^{-5}$ but only at $A_{\rm V}=0.1$. If the hydrogenation is included,  the peak of C$_{60}$ abundance shifts to $A_{\rm V}=0.9$, and a trough in abundance of PAHs and other related molecules appears near $A_{\rm V}=0.1$. The observed profiles of PAH and fullerene emission are close to the modelled ones with the hydrogenation, although precise juxtaposition requires further investigations. The modelled $f_{\rm C}^{{\rm C}_{60}}$ is consistent with observational estimates, although the modelled ratio $I_{11.2}/I_{18.9}$ is lower than the observed one. Based on the results of our calculations, we conclude that fullerenes indeed can be formed from PAHs through photo-dissociation, but only in locations that are close to the ionising source, while fullerenes at other locations must have been transferred there or have another origin. 

In NGC~7023, if hydrogenation is considered, the modelled fraction of carbon in C$_{60}$ and C$_{70}$ reaches the value of $10^{-2}$ at the distance of $5\arcsec$ and become around $10^{-5}$ and $10^{-7}$, correspondingly, at the distance of $25\arcsec$. Value of $f_{\rm C}^{\rm PAHs}$ changes from $10^{-8}$ and $10^{-5}$ along the PDR. The observations and results of the modelling with hydrogenation are consistent generally with each other, and we conclude that conditions are favourable for fullerene formation from PAHs within the considered region of the object.  

We analysed the results obtained with modelling on the grid of parameters and found general trends in relations between $f_{\rm C}^{{\rm C}_{60}}$, $f_{\rm C}^{{\rm C}_{70}}$ and $f_{\rm C}^{{\rm PAHs}}$ and the parameter $U T_{\rm gas} n({\rm H})$. The maximum of $f_{\rm C}^{{\rm PAHs}}$ is reached at the lowest value of the parameter, while the maxima of $f_{\rm C}^{{\rm C}_{60}}$ and $f_{\rm C}^{{\rm C}_{70}}$ are reached when $U T_{\rm gas} n({\rm H})$ is between $\approx10^{4}$ and $10^{10}$~K~cm$^{-3}$. At these conditions $f_{\rm C}^{{\rm C}_{60}}$ and $f_{\rm C}^{{\rm C}_{70}}$ are comparable by value. $f_{\rm C}^{{\rm C}_{60}}$ and $f_{\rm C}^{{\rm C}_{70}}$ decrease, when the parameter is higher than $\approx10^{10}-10^{12}$~K~cm$^{-3}$. At some conditions C$_{58}$ and C$_{60}$H$_{x}$ (fulleranes) may appear. Their maxima of $f_{\rm C}$ are $\approx10^{-6}$ and $\approx10^{-10}$, correspondingly, which are lower than the maximum of $f_{\rm C}^{{\rm C}_{60}}$ by 4 and 6 orders of magnitude, correspondingly.  

We conclude that the process of fullerene formation through PAH dissociation is likely to take place in PDRs and the ISM. However, its efficiency is strongly sensitive to the physical conditions (radiation field intensity, gas temperature, hydrogen and electronic density). The object's age is also an important factor as the abundance of fullerenes changes with time. Contribution of fullerenes formed via the top-down scenario cannot explain the observations in the Orion Bar, and there might be another alternative source of fullerenes in the ISM or they should be transferred efficiently. PAH dissociation can provide an efficient source of fullerenes in the closest surroundings of bright UV radiation sources. Hydrogenated fullerenes (fulleranes) are of low abundance in the ISM under the conditions considered in this work, and can be hardly detected by observations. 

\section*{Acknowledgements}

We thank the referee for reading the paper and providing valuable comments. The work was supported by the Russian Science Foundation (project 18-13-00269).

\section*{Data availability}

The modelling results presented in this paper are available on request.

\bibliographystyle{mnras} 
\bibliography{fulleren}

\bsp	
\label{lastpage}
\end{document}